\documentclass{elsart}
\usepackage{graphicx}% Include figure files
\usepackage{dcolumn}% Align table columns on decimal point
\usepackage{bm}% bold math
\usepackage{color}
\usepackage{latexsym}
\journal{Nuclear Physics A}
\usepackage{latexsym}
\begin{document}
%%%%%%%%%%%%%%%%%%%%%%%%%%%%%%%%%%%%%%%%%%%%%%%%%%%%%%%%%%%%%%%%%%
\begin{frontmatter}
\title{Relativistic nuclear model with point-couplings
constrained by QCD and chiral symmetry\thanksref{now}}
\thanks[now]{Work supported
in part by BMBF, DFG and INFN}
\author[Pao,Wei]{P. Finelli},
\author[Kai]{N. Kaiser},
\author[Vre]{D. Vretenar},
\author[Wei,Kai]{W. Weise}
\address[Pao]{Physics Department, University of Bologna, and INFN - Bologna,
I-40126 Bologna, Italy}
\address[Kai]{Physik-Department, Technische Universit\"at M\"unchen,
D-85747 Garching, Germany}                                                    
\address[Vre]{Physics Department, Faculty of Science, University of
Zagreb, 10 000 Zagreb, Croatia}
\address[Wei]{ECT$^*$, I-38050 Villazzano (Trento), Italy
}
\begin{abstract}
We derive a microscopic relativistic point-coupling model of
nuclear many-body dynamics constrained by
in-medium QCD sum rules and chiral symmetry.
The effective Lagrangian is characterized by density
dependent coupling strengths, determined by
chiral one- and two-pion exchange and by QCD sum rule constraints
for the large isoscalar nucleon self-energies that arise through
changes of the quark condensate and the quark density
at finite baryon density. This approach is tested in the analysis
of the equations of state for symmetric and asymmetric nuclear matter,
and of bulk and single-nucleon properties of finite nuclei.
In comparison with purely phenomenological mean-field approaches,
the built-in QCD constraints and the explicit treatment
of pion exchange restrict the freedom in adjusting
parameters and functional forms of density dependent couplings.
It is shown that chiral (two-pion exchange) fluctuations play
a prominent role for nuclear binding and
saturation, whereas strong scalar and vector fields of
about equal magnitude and
opposite sign, induced by changes of the QCD vacuum in the presence
of baryonic matter, generate the large effective
spin-orbit potential in finite nuclei.
\end{abstract}  
\date{\today}
\end{frontmatter}

%=========================================================================
%=========================================================================
%  Section 1

\section{\label{secI}Introduction}
%=========================================================================

Concepts of effective field theory and
density functional theory are at the basis of many
successful nuclear structure models. Quantum Hadrodynamics (QHD),
in particular, presents a field theoretical
framework of Lorentz-covariant, meson-nucleon or point-coupling
models of nuclear dynamics, that have been successfully
employed in studies of a variety of nuclear phenomena~\cite{SW.97}.
Structure models based on the relativistic mean-field (RMF)
approximation of QHD have been applied to describe
spherical and deformed nuclei
all over the periodic table~\cite{Rin.96}. More recently,
this relativistic framework has also been used in studies of
the structure of exotic nuclei with extreme isospin
and close to the particle drip lines.

The effective Lagrangians of QHD consist of known long-range interactions
constrained by symmetries
and a set of generic short-range interactions. 
The most successful QHD models are
phenomenological, with parameters adjusted to reproduce
the nuclear matter equation of state and a set of global properties
of spherical closed-shell nuclei. On the other hand, the
description of nuclear many-body dynamics should ultimately be
constrained by the underlying theory of the strong interactions -
Quantum Chromodynamics (QCD). It is desirable
to have a more explicit connection between low-energy QCD and nuclear
phenomenology.

As pointed out in Refs.~\cite{SW.97,FS.00,FST},
the effective hadronic Lagrangians of QHD and related models
are consistent with the symmetries of QCD:
Lorentz invariance, parity invariance, electromagnetic gauge invariance,
isospin and chiral symmetry. However, QHD calculations do not include
pions explicitly.
Effects of two-pion exchange are treated implicitly through
a phenomenological scalar, isoscalar mean-field.
Using the linear $\sigma$-model Lagrangian to construct a
dynamical model for the $s$-wave $\pi\pi$ scattering amplitude,
it has been shown that the exchange of a light scalar boson can
be viewed as an approximate description of the exchange of
correlated $s$-wave pion pairs~\cite{Lin}.
While QCD symmetries constrain
the effective QHD Lagrangians by restricting the form of possible
interaction terms~\cite{FST}, the empirical data set of bulk
and single-particle properties of finite nuclei can only determine
six or seven parameters in the general expansion of the effective
Lagrangian in powers of the fields
and their derivatives~\cite{FS.00b}. Such a general expansion is
guided by the ``naive dimensional analysis"
(NDA)~\cite{Man83,FML.96,Rusnak:1997dj,Burvenich:2001rh}. NDA tests the
coefficients of the expansion for ``naturalness", i.e. this method permits
to control the orders of magnitude of the coupling constants. NDA can
exclude some interaction terms because their couplings would be
``unnatural", but it cannot determine the model parameters at the
level of accuracy required for a quantitative analysis of
nuclear structure data.

The success
of the relativistic nuclear mean-field phenomenology
has been attributed to the large Lorentz scalar and four-vector nucleon
self-energies, which are at the basis of all QHD models~\cite{FS.00}.
Even though the presence of large scalar and vector potentials cannot
be directly verified experimentally, there is strong evidence, in particular
from nuclear matter saturation and spin-orbit splittings in finite
nuclei, that the magnitude of each of these potentials is of the 
order of several hundred MeV in the nuclear interior. 
Studies based on finite-density QCD sum rules have shown how 
such large scalar and vector nucleon self-energies arise naturally 
from in-medium changes of the scalar quark condensate 
and the quark density~\cite{CFG.91,DL.90,Jin.94}.
In applications of QHD models to finite nuclei, however, the
scalar and vector nucleon self-energies are treated in a purely
phenomenological way. In some models they arise from the exchange
of ``sigma'' and ``omega'' mesons,
with adjustable strength parameters determined by the nuclear matter
saturation and nuclear structure data. Additional nonlinear terms,
or density-dependent coupling constants
adjusted to Dirac-Brueckner self-energies in nuclear matter,
are necessary in order to obtain a satisfactory nuclear matter
equation of state and to reproduce the empirical bulk properties
of finite nuclei.

The approach we adopt in the present work is based
on the following conjectures that are believed to establish
links between QCD, its symmetry breaking
pattern, and the nuclear many-body problem:
\begin{itemize}
\item
The nuclear ground state is characterized by large scalar and vector
fields of approximately equal magnitude and opposite sign.
These fields have their
origin in the in-medium changes of the scalar quark condensate
(the chiral condensate) and of the quark density.\\
\item
Nuclear binding and saturation arise prominently
from chiral (pionic) fluctuations (reminiscent of van der Waals forces)
in combination with Pauli blocking effects,
superimposed on the condensate background fields and calculated
according to the rules of in-medium chiral perturbation theory.\\
\end{itemize}

The first item has a direct connection with QCD sum rules at finite density.
The second item is motivated
by the observation that, at nuclear matter saturation density,
the nucleon Fermi momentum $k_f$ and the pion mass $m_\pi$ represent
comparable scales. 
We therefore include pions as {\it explicit} degrees of freedom
in the description of nuclear many-body dynamics. 
Hammer and Furnsthal~\cite{Ham_Fur} pointed out earlier
that an effective field theory of dilute Fermi systems
must include pions for a proper description of the long-range physics.
We support this statement by explicit calculations.
While it is well known
that the lowest order Fock terms from pion exchange are relatively small,
the primary importance of chiral dynamics in nuclear systems is
manifest in two-pion exchange processes combined with
Pauli blocking effects which lead beyond standard
mean-field approximations. 
\par The broad spectral distribution
of intermediate $\pi-\pi$ pairs with $J^P=0^+$, exchanged between nucleons,
has large low-mass components starting at $2 m_\pi$.
Hence their propagation in the presence of the nuclear Fermi
sea is {\it resolved} at the Fermi momenta $k_f \simeq 2 m_\pi$
characteristic of nuclear matter. Guided by such considerations
the nuclear matter problem has been investigated
in Ref.~\cite{KFW.01} within the framework of chiral effective field theory.
The calculations have been performed using in-medium chiral perturbation theory
to three-loop order and incorporate the one-pion exchange Fock term, iterated
one-pion exchange and irreducible two-pion exchange. The resulting nuclear
matter equation of state is expressed as an expansion in powers of the Fermi
momentum $k_f$. The expansion coefficients are functions of $k_f/m_\pi$, the
dimensionless ratio of the two relevant small scales. 
The calculation involves one single momentum space 
cutoff $\Lambda$ which encodes NN-dynamics at short
distances. This cutoff scale is the only free parameter. Its value can be
adjusted to the energy per particle $\bar E(k_{f0})$
at saturation density. With  $\Lambda
\simeq  0.65$ GeV adjusted to $\bar{E}(k_{f0}) = -15.3$ MeV,
the calculated equation of state gives the saturation density  $\rho_0=0.178
$\,fm$^{-3}$, the compression modulus $K_0 = 255$ MeV, and the asymmetry energy
$A(k_{f0}) = 33.8$\,MeV at saturation density. 
The corresponding momentum- and density-dependent 
single-particle potential $U(p,k_{f})$ of nucleons in 
isospin-symmetric nuclear matter has been calculated in Ref.~\cite{KFW.02}.
In particular, it has been shown that, at equilibrium nuclear matter density,
chiral one- and two-pion exchanges generate an attractive nuclear mean-field
for nucleons at the bottom of the Fermi sea, 
with a depth of $U(0,k_{f0}) = -53.2$ MeV, in good agreement 
with the depth of the empirical optical-model
potential deduced by extrapolation from elastic nucleon-nucleus
scattering data, and with the average nuclear potentials employed
in shell-model and mean-field calculations for finite nuclei.

In Ref.~\cite{Finelli:2002na} we have outlined an approach to the
nuclear many-body problem, both for nuclear
matter and finite nuclei, which emphasizes the connection with
the condensate structure of the QCD ground state and 
pion dynamics based on spontaneous chiral symmetry breaking. 
A relativistic point coupling model has been used, with density dependent  
couplings governed by scales of low-energy QCD. We have shown that
chiral fluctuations play a prominent role for nuclear binding and
saturation, whereas strong scalar and vector fields of equal magnitude and
opposite sign, induced by changes of the QCD vacuum in the presence
of baryonic matter, drive the large spin-orbit splitting in finite nuclei.
The magnitude of these condensate background nucleon self-energies
has been determined by the leading order in-medium QCD sum rules.

In a different implementation, a model emphasizing the same connections
to QCD symmetries and structure, was introduced and applied at the
one-baryon-loop level to finite nuclei and nuclear matter in Ref.~\cite{FTS}.
 
In this work we expand our previous studies and
present a generalized and consistent version of the
relativistic point-coupling model introduced in Ref.~\cite{Finelli:2002na}.
For instance, in the previous work following Ref.~\cite{KFW.01},
it was still assumed that  
at nuclear matter saturation density the ratio of
the scalar and vector condensate potentials, arising through
changes in the quark condensate and the quark density
at finite baryon density, equals $-1$ and that the nuclear
binding  and saturation result entirely from pionic (chiral)
fluctuations. This approximation is relaxed in the present generalized approach
and the effects of the condensate background fields at finite density
are included already at the nuclear matter level. Eventually we
will demonstrate that with a small number of parameters, determined
by the underlying microscopic dynamics,
and with minor fine-tuning of higher order corrections,
it is possible to describe nuclear matter,
as well as properties of finite nuclei, on a level comparable to
phenomenological relativistic mean-field models.
  
In Sec.~\ref{secII} we present
the formalism and develop the relativistic point-coupling model
constrained by QCD and chiral symmetry. An effective interaction,
characterized by density dependent coupling parameters of the
local four-point interaction terms, is derived in Sec.~\ref{secIII}
and adjusted to the equation of state for symmetric and asymmetric
nuclear matter. In Sec.~\ref{secIV} the model is employed in
self-consistent calculations of ground-state properties of
light and medium-heavy $N\approx Z$ nuclei. The calculated bulk properties and
single-nucleon spectra are compared with experimental data and
with results of the standard boson-exchange relativistic mean-field model.
Summary and conclusions are presented in Sec.~\ref{secV}.

%=========================================================================
%  Section 2
\section{\label{secII}Point-coupling model}
%=========================================================================
\subsection{Lagrangian}
%=========================================================================

The relativistic point-coupling Lagrangian is
built from basic densities and currents bilinear in the Dirac
spinor field $\psi$ of the nucleon:
\begin{equation}
  {\bar{\psi}} {\mathcal O}_\tau \Gamma  {\psi}
  \quad,\quad
  {\mathcal O}_\tau\in\{ {1},\tau_i\}
  \quad,\quad
  \Gamma\in\{1,\gamma_\mu,\gamma_5,\gamma_5\gamma_\mu,\sigma_{\mu\nu}\}\; .
\end{equation}
Here $\tau_i$ are the isospin Pauli matrices and
$\Gamma$ generically denotes the Dirac matrices.
The interaction terms of the Lagrangian are products of these
bilinears. In principle, a general effective Lagrangian
can be written as a power series in the currents 
${\bar{\psi}} {\mathcal O}_\tau \Gamma  {\psi}$ and their derivatives.
However, it is well known from numerous applications
of relativistic mean-field models that properties of symmetric and asymmetric
nuclear matter, as well as empirical ground state properties of finite
nuclei, constrain only the isoscalar-scalar, the isoscalar-vector,
the isovector-vector, and to a certain extent the isovector-scalar
channels.
\par
The model that we consider in the  
present work includes the following four-fermion interaction vertices:
\begin{center}
\begin{tabular}{ll}
 isoscalar-scalar:   &   $(\bar\psi\psi)^2$\\
 isoscalar-vector:   &   $(\bar\psi\gamma_\mu\psi)(\bar\psi\gamma^\mu\psi)$\\
 isovector-scalar:  &  $(\bar\psi\vec\tau\psi)\cdot(\bar\psi\vec\tau\psi)$\\
 isovector-vector:   &   $(\bar\psi\vec\tau\gamma_\mu\psi)
                         \cdot(\bar\psi\vec\tau\gamma^\mu\psi)$ .\\
\end{tabular}
\end{center}
Vectors in isospin space are denoted by arrows.
The model is defined by the Lagrangian density
\begin{equation}
\mathcal{L} = \mathcal{L}_{\rm free} + \mathcal{L}_{\rm 4f}
  + \mathcal{L}_{\rm der} + \mathcal{L}_{\rm em},
\label{Lag}
\end{equation}
with the four terms specified as follows:
\begin{eqnarray}
\label{Lag2}
\mathcal{L}_{\rm free} & = &\bar{\psi}
   (i\gamma_{\mu}\partial^{\mu} -M)\psi \; ,\\
\label{Lag3}
\mathcal{L}_{\rm 4f} & = &
   - \frac{1}{2}~G_{S}(\hat{\rho}) (\bar{\psi}\psi)(\bar{\psi}\psi) \nonumber\\
   & ~ & -\frac{1}{2}~G_{V}(\hat{\rho})(\bar{\psi}\gamma_{\mu}\psi)
   (\bar{\psi}\gamma^{\mu}\psi) \nonumber\\
   & ~ & - \frac{1}{2}~G_{TS}(\hat{\rho})
   (\bar{\psi}\vec{\tau}\psi) \cdot (\bar{\psi} \vec{\tau} \psi) \nonumber\\
   & ~ & - \frac{1}{2}~G_{TV}(\hat{\rho})(\bar{\psi}\vec{\tau}
   \gamma_{\mu}\psi)\cdot (\bar{\psi}\vec{\tau} \gamma^{\mu}\psi) \; ,\\
\label{Lag4}
\mathcal{L}_{\rm der} & = & -\frac{1}{2}~D_{S}(\hat{\rho}) (\partial_{\nu}
   \bar{\psi}\psi)(\partial^{\nu}\bar{\psi}\psi)\nonumber\\
 & ~ & - \frac{1}{2}~D_{V}(\hat{\rho})
   (\partial_{\nu}\bar{\psi}\gamma_{\mu}
   \psi)(\partial^{\nu}\bar{\psi}\gamma^{\mu}\psi) + \; \ldots~,\\
\label{Lag5}
\mathcal{L}_{\rm em} & = &
   eA^{\mu}\bar{\psi}\frac{1+\tau_3}{2}\gamma_{\mu}\psi
   -\frac{1}{4} F_{\mu\nu}F^{\mu\nu} \; .
\end{eqnarray}
This Lagrangian (\ref{Lag})-(\ref{Lag5}) is understood  to be formally used
in the mean-field approximation~\cite{FLW.95}, with fluctuations encoded
in density-dependent  couplings $G_i(\hat{\rho})$ and $D_i(\hat{\rho})$,
to be specified in detail later.
In addition to the free nucleon Lagrangian $\mathcal{L}_{\rm free}$
and the interaction terms contained in
$\mathcal{L}_{\rm 4f}$, when applied to finite nuclei the model
must include the coupling $\mathcal{L}_{\rm em}$
of the protons to the electromagnetic field $A^\mu$,
and derivative terms contained in $\mathcal{L}_{\rm der}$\footnote{
In the terms $\partial_\nu(\bar\psi \Gamma \psi)$ the derivative is
understood to act on both $\bar\psi$ and $\psi$
}. 
One can, of course, construct many more derivative terms
of higher orders. In this way we take into
account leading effects of finite range
interactions that are important for a quantitative fine-tuning
of nuclear properties.  
In practice it will turn out that a single term
$D_{S}(\partial_{\nu} \bar{\psi}\psi)(\partial^{\nu}\bar{\psi}\psi)$ 
is already sufficient to account
for the detailed surface structure of the nuclei considered.

Our model Lagrangian is formally equivalent to the ones used in
the standard relativistic mean-field point-coupling models of
Refs.~\cite{Manakos:wu,Mad,FML.96,Rusnak:1997dj,Burvenich:2001rh}.
The underlying dynamics is, however, quite different.
The parameters of the interaction terms in the standard
point-coupling models have constant values adjusted to reproduce
the nuclear matter equation of state and a set of global
properties of spherical closed-shell nuclei. In order to
describe properties of finite nuclei on a quantitative level,
these models include also some higher order interaction terms,
such as six-fermion vertices $(\bar\psi\psi)^3$, and
eight-fermion vertices $(\bar\psi\psi)^4$ and
$[(\bar\psi\gamma_\mu\psi)(\bar\psi\gamma^\mu\psi)]^2$.
The model that we develop in this work includes only
second order interaction terms. On the other hand, the
coupling parameters are not constants, but rather functions
of the nucleon density operator $\hat{\rho}$. Their functional dependence
will be determined from finite-density QCD sum rules and
in-medium chiral perturbation theory.

Medium dependent vertex functions have also been considered
in the framework of relativistic mean-field meson-exchange
models. In the density dependent relativistic hadron field (DDRH)
models the medium dependence of the meson-nucleon vertices is
expressed as a functional of the baryon field operators.
The meson-nucleon vertex functions are determined either
by mapping the nuclear matter Dirac-Brueckner nucleon self-energies
in the local density approximation~~\cite{FLW.95,JL.98,HKL.01}, or
the parameters of an assumed phenomenological density dependence of
the meson-nucleon couplings are adjusted
to reproduce properties of symmetric and asymmetric nuclear matter
and finite nuclei~\cite{TW.99,Niksic:2002yp}. In practical applications
of the DDRH models the meson-nucleon couplings are assumed to be
functions of the baryon density $\psi^\dagger \psi$.
In a relativistic framework the couplings can also depend on
the scalar density $\bar{\psi} \psi$. 
Nevertheless, expanding in $\psi^\dagger \psi$ is the natural 
choice, for several reasons. The baryon density is connected
to the conserved baryon number, unlike the scalar density
for which no conservation law exists. The scalar density is
a dynamical quantity, to be determined self-consistently
by the equations of motion, and expandable in powers of the Fermi momentum.
For the meson-exchange models it has been shown that the
dependence on baryon density alone provides a more direct relation
between the self-energies of the density-dependent
hadron field theory and the Dirac-Brueckner microscopic
self-energies~\cite{HKL.01}. 
Moreover, the pion-exchange contributions to the nucleon self-energy,
as calculated using in-medium chiral perturbation theory, are directly
given as expansions in powers of the Fermi momentum. 
Following these considerations, we
express the coupling strengths of the interaction terms
as functions of the baryon density, represented by the operator
$\hat{\rho}=\psi^\dagger \psi$ in the rest frame of the many-body system.
\\ \bigskip \bigskip

%%%%%%%%%%%%%%%%%%%%%%%%%%%%%%%%%%%%%%%%%%%%%%%%%%%%%%%%%%%%%%%%%%%%%%%%%%%%

\subsection{Equation of motion and nucleon self-energies}

%%%%%%%%%%%%%%%%%%%%%%%%%%%%%%%%%%%%%%%%%%%%%%%%%%%%%%%%%%%%%%%%%%%%%%%%%%%%

The single-nucleon Dirac equation is derived by variation of the
Lagrangian (\ref{Lag}) with respect to $\bar{\psi}$:
\begin{equation}
\label{Dirac}
   [\gamma_{\mu}(i\partial^{\mu} - V^\mu) -
   (M + S) ]\psi = 0\; ,
\end{equation}
where
\begin{equation}
   V^\mu = \Sigma^{\mu} +
   \vec{\tau} \cdot \vec{\Sigma}^{\mu}_{T}
   +\Sigma_{r}^{\mu}
\end{equation}
and
\begin{equation}
   S = \Sigma_S + \vec{\tau} \cdot \vec{\Sigma}_{TS} + \Sigma_{rS} \; ,
\end{equation}
with the nucleon self-energies defined by the following relations
\begin{eqnarray}
\label{self1}
\Sigma^{\mu} & = & G_V (\bar{\psi} \gamma^\mu \psi) -
   eA^{\mu}\frac{1+\tau_3}{2} \; ,\\
\label{self2}
   \vec{\Sigma}^{\mu}_{T} & = & G_{TV} (\bar{\psi} 
   \vec{\tau} \gamma^\mu \psi) \; ,\\
\label{self3}
   \Sigma_S & = & G_S (\bar{\psi} \psi) - D_S \Box (\bar{\psi} \psi) \; ,\\
\label{self4}
   \vec{\Sigma}_{TS} & = & G_{TS} (\bar{\psi} \vec{\tau} \psi)\; ,\\
\label{self5}
   \Sigma_{rS} &  = & - \frac{\partial D_S}{\partial \hat{\rho}} 
   (\partial_{\nu} j^{\mu}) u_{\mu} (\partial^{\nu} (\bar{\psi} \psi)) \; ,\\
\label{self8}
   \Sigma_r^{\mu} & = & 
   \frac{u^\mu}{2} \left( 
   \frac{\partial G_S}{\partial \hat{\rho}} (\bar{\psi} \psi)
   (\bar{\psi} \psi) +
   \frac{\partial G_{TS}}{\partial \hat{\rho}} 
   (\bar{\psi} \vec{\tau} \psi) \cdot
   (\bar{\psi} \vec{\tau} \psi) \right. \nonumber\\
   & ~ &  
   + \frac{\partial G_V}{\partial \hat{\rho}}
   (\bar{\psi} \gamma^\mu \psi) (\bar{\psi} \gamma_\mu \psi)
   + \frac{\partial G_{TV}}
   {\partial \hat{\rho}} (\bar{\psi} \vec{\tau} \gamma^\mu \psi) \cdot
   (\bar{\psi} \vec{\tau} \gamma_\mu \psi) \nonumber\\
 & ~ & \left. +  
   \frac{\partial D_S}{\partial \hat{\rho}} (\partial^{\nu} (\bar{\psi} \psi))
   (\partial_{\nu} (\bar{\psi} \psi)) \right)
   \; .
\end{eqnarray}
We write $\hat{\rho} u^{\mu} = \bar{\psi} \gamma^\mu \psi$
and the four-velocity $u^{\mu}$ is defined
as $(1-{\bm v}^2)^{-1/2}(1,{\bm v})$
where $\bm v$ is the three-velocity vector ($\bm v=0$
in the rest-frame of the nuclear system).
In addition to the isoscalar-vector $\Sigma^{\mu}$,
isoscalar-scalar $\Sigma_S$, isovector-vector $\vec{\Sigma}^{\mu}_{T}$
and isovector-scalar $\vec{\Sigma}_{TS}$
self-energies, the density dependence of the vertex functions
produces the {\it rearrangement} contributions $\Sigma_{rS}$ and 
$\Sigma_r^{\mu}$.
The rearrangement terms result from the variation of the vertex
functionals with respect to the baryon fields in the density 
operator $\hat{\rho}$
(which coincides with the baryon density in the nuclear matter rest-frame).
For a model with density dependent couplings, the inclusion
of the rearrangement self-energies
is essential for energy-momentum conservation and thermodynamical consistency  
(i.e. for the pressure equation derived from the thermodynamic definition and 
from the energy-momentum tensor)~\cite{FLW.95,TW.99}.

The density dependence of the coupling strengths which determine 
the self-energies (\ref{self1}-\ref{self8}), 
will be constrained by in-medium QCD sum rules
and chiral pion dynamics, to be specified in detail in the following
subsections~\ref{secIII-2} and~\ref{secIII-3}.
\par
When applied to nuclear matter or ground-state properties of
finite nuclei, the point-coupling model is understood to be used in the
mean-field approximation.
The ground state of a nucleus with A nucleons
is the antisymmetrized product of the lowest occupied single-nucleon
self-consistent stationary solutions of the Dirac equation Eq.~(\ref{Dirac}).
The ground state energy is the sum of the single-nucleon energies plus
a functional of proton and neutron scalar and particle densities.
%
%=========================================================================
%  Section 3
\section{\label{secIII}Nuclear matter equation of state}
%=========================================================================
%
\subsection{Framework}
In the translationally invariant infinite nuclear matter
all terms involving the derivative couplings (\ref{Lag4}) drop out.
The spatial components of the four-currents vanish,
and the densities are calculated by taking expectation values
\begin{eqnarray}
   \rho_s & = & \langle \Phi|\bar{\psi} \psi|\Phi \rangle 
    = \rho_s^p + \rho_s^n \; , \\
   \rho & = & \langle \Phi|\bar{\psi}\gamma^{0} \psi|\Phi \rangle 
    = \rho^p + \rho^n \; , \\
   \rho_{s3} & = & \langle \Phi|\bar{\psi} \tau_3 \psi|\Phi \rangle 
    = \rho_s^p - \rho_s^n \; , \\
   \rho_{3} & = & \langle \Phi|\bar{\psi} \tau_3 \gamma^{0} 
   \psi|\Phi \rangle = \rho^p - \rho^n  \; ,
\end{eqnarray}
where $|\Phi \rangle$ is the nuclear matter ground state.
The energy density $\epsilon$ and the pressure $P$ are derived from the
energy-momentum tensor as
\begin{eqnarray}
   \epsilon & = & \epsilon_{kin}^n + \epsilon_{kin}^p
   - \frac{1}{2}G_S \rho_s^2 - \frac{1}{2}G_{TS} \rho_{s3}^2
   + \frac{1}{2}G_V \rho^2 + \frac{1}{2}G_{TV} \rho_{3}^2 \; ,\label{epsilon}\\
 & ~ & \nonumber \\
 P & = & \tilde{E}^p \rho^p + \tilde{E}^n \rho^n - \epsilon_{kin}^p -
   \epsilon_{kin}^n
   + \frac{1}{2} G_V \rho^2 + \frac{1}{2} G_{TV}\rho_{3}^2  
   + \frac{1}{2} G_S \rho_s^2 + \frac{1}{2} G_{TS} \rho_{s3}^2 \nonumber \\
 & ~ &
   + \frac{1}{2} \frac{\partial G_S} {\partial \rho} \rho_s^2 \rho +
   \frac{1}{2} \frac{\partial G_V}{\partial \rho} \rho^3 +
   \frac{1}{2} \frac{\partial G_{TV}}{\partial \rho} \rho_{3}^2
   \rho + \frac{1}{2} \frac{\partial G_{TS}} {\partial \rho}
   \rho_{s3}^2 \rho \label{pressure} \; .
\end{eqnarray}
The particle density $\rho^i$ is related to the 
Fermi momentum $k_f^i$ in the usual way,
\begin{equation}
   \rho^i = \frac{2}{(2\pi)^3} \int\limits_{|\vec{k}|\le k^i_f} d^3 k  =
   \frac{(k^i_f)^3}{3\pi^2} \; ,
\end{equation}
where the index $i=p,n$ refers to protons and neutrons, respectively.
The corresponding scalar densities are determined by the self-consistency
relation
\begin{equation}
   \rho_s^i = \frac{2}{(2\pi)^3} \int\limits_{|\vec{k}|\le k^i_f} d^3 k
   \frac{M^*_i}{\sqrt{\vec{k}^2
   + (M^*_i)^2}} = \frac{M^*_i}{2\pi^2} \left[ k^i_f \tilde{E}^i - (M^*_i)^2
   \ln \frac{k^i_f + \tilde{E}^i}{M^*_i} \right] \; ,
\end{equation}
with 
\begin{equation}
   \tilde{E}^i = \sqrt{(k^i_f)^2 + (M^*_i)^2} \; ,
\end{equation}
and the effective nucleon masses
\begin{eqnarray}
   M^*_{p,n} &=& M + G_S \rho_s \pm G_{TS} \rho_{s3} \; .
\end{eqnarray}
The kinetic contributions to the energies of the protons and neutrons in
nuclear matter are calculated from
\begin{equation}
\epsilon^i_{kin} = \frac{2}{(2\pi)^3} \int\limits_{|\vec{k}|\le k^i_f}
   d^3 k \sqrt{\vec{k}^2 + (M^*_i)^2} =
   \frac{1}{4} [ 3 \tilde{E}^i \rho^i + M^*_i \rho_s^i ] \; , \quad i=p,n \; .
\end{equation}
Note that, in contrast to the energy density, {\it rearrangement} contributions
appear explicitly in the expression for the pressure. 

%%%%%%%%%%%%%%%%%%%%%%%%%%%%%%%%%%%%%%%%%%%%%%%%%%%%%%%%%%%%%%%%%%%%%%%%%%%%%

\subsection{\label{secIII-2}QCD constraints}

%%%%%%%%%%%%%%%%%%%%%%%%%%%%%%%%%%%%%%%%%%%%%%%%%%%%%%%%%%%%%%%%%%%%%%%%%%%%

We proceed now with a central theme of our work: establishing connections
between the density-dependent point couplings in
(\ref{self1}-\ref{self8}) and constraints
from QCD. Two key features of low-energy, non perturbative QCD are at 
the origin of this discussion: 
the presence of a non-trivial vacuum characterized by strong
condensates and the important role  of pionic fluctuations governing
the low-energy, long wavelength dynamics according to the rules imposed by
spontaneously broken chiral symmetry.

The basic conjecture underlying our model is, consequently,
that the nucleon isoscalar self-energies arise primarily
through changes in the quark condensate and in the quark density at
finite baryon density, together with chiral (pionic) fluctuations
induced by one- and two-pion exchange interactions. In-medium QCD sum rules
relate the changes of the scalar quark condensate and the quark density at
finite baryon density, with the scalar and vector self-energies
of a nucleon in the nuclear medium. In leading order which should be valid at
densities below and around saturated nuclear matter, the condensate
part of the scalar self-energy is expressed in terms of the density dependent
chiral condensate as follows~\cite{CFG.91,DL.90,Jin.94}:
\begin{equation}
  \Sigma^{(0)}_S = - \frac{8 \pi^2}{\Lambda_B^2} [
  \langle \bar{q} q \rangle_\rho - \langle \bar{q} q \rangle_0 ]
  = - \frac{8 \pi^2}{\Lambda_B^2}~\frac{\sigma_N}{m_u +m_d} \rho_s \; .
\end{equation}
The chiral vacuum condensate $\langle \bar{q} q \rangle_0$ is a measure
of spontaneous chiral symmetry breaking in QCD. At a renormalization scale
of about 1 GeV (with quark masses $m_u + m_d \simeq 12$ MeV) 
its value \cite{Pic} is $\langle \bar{q} q \rangle_0 \simeq 
- (240~{\rm MeV})^3 \simeq - 1.8~{\rm fm}^{-3}$.
In order to appreciate the strength of this scalar condensate, 
note that its magnitude is more than a factor of ten larger 
than the baryon density in the bulk of a heavy nucleus.
The difference between the vacuum condensate $\langle \bar{q} q \rangle_0$
and the one at finite density involves the nucleon sigma term,
$\sigma_N = \langle N| m_q \bar{q} q |N \rangle$, to this order.
Furthermore, $\Lambda_B \approx 1$GeV is a characteristic scale 
(the Borel mass) roughly separating perturbative and non-perturbative 
domains in the QCD sum rule analysis. To the same order in the 
condensates with lowest dimension, the time component of the 
vector self-energy is
\begin{equation}
  \Sigma_V^{(0)} = \frac{64 \pi^2}{3 \Lambda_B^2} 
  \langle q^\dagger q \rangle_\rho
  = \frac{32 \pi^2}{\Lambda_B^2} \rho \; ,
\end{equation}
where the quark baryon density is simply related to that of the nucleons by
$\langle q^\dagger q \rangle_\rho = \frac{3}{2} \rho$.

When taken at the same leading order, the QCD sum rule analysis also identifies
the free nucleon mass at $\rho=0$ according to Ioffe's formula~\cite{Iof},
$M = - \frac{8\pi^2}{\Lambda_B^2} \langle \bar{q} q \rangle_0$.
Together with the Gell-Mann, Oakes, Renner relation\\ $(m_u +m_d)
\langle \bar{q} q \rangle_0 = - m_\pi^2 f_\pi^2$
(where $m_\pi$ is the pion mass and $f_\pi = 92.4$ MeV is
the pion decay constant), one finds
\begin{equation}
\label{back1}
   \Sigma_S^{(0)} (\rho) = M^*(\rho) -M =
   - \frac{\sigma_N M}{m_\pi^2 f_\pi^2} \rho_s
\end{equation}
and
\begin{equation}
\label{back2}
   \Sigma_V^{(0)} (\rho) = \frac{4 (m_u + m_d)M}{m_\pi^2f_\pi^2} \rho \; .
\end{equation}
\par
Given these self-energies arising from the condensate background,
the corresponding equivalent point-coupling strengths $G_{S,V}^{(0)}$
are simply determined by
\begin{equation}
\label{sigmaS}
   \Sigma_S^{(0)} = G_S^{(0)} \rho_s
\end{equation}
and
\begin{equation}
\label{sigmaV}
   \Sigma_V^{(0)} = G_V^{(0)} \rho \; .
\end{equation}
The second important ingredient is the contribution to the nucleon 
self-energies from chiral fluctuations related to pion-exchange processes
(primarily iterated one-pion exchange 
with small corrections from one-pion exchange Fock terms).
While pion-exchange processes are not explicitly
covered by standard Hartree-type mean field
descriptions, their effects can nevertheless be
re-expressed as density dependent corrections to the mean fields.
The corresponding point-coupling strengths, $G^{(\pi)}_{S,V}(\rho)$ etc.,
and their explicit density dependencies are calculated using in-medium
chiral perturbation theory to a given loop order.

Combining effects from leading QCD condensates and pionic fluctuations,
the strength parameters of the isoscalar four-fermion
interaction terms in the Lagrangian~(\ref{Lag}) are:
\begin{equation}
\label{couplings}
   G_{S,V}(\rho) = G_{S,V}^{(0)} + G_{S,V}^{(\pi)} (\rho)\; .
\end{equation}
For the isovector channels we assume in this work that only pionic
(chiral) fluctuations contribute.

In Ref.~\cite{KFW.01} the nuclear
matter equation of state (EOS) has been calculated using in-medium
chiral perturbation theory up to three loop order in the energy density,
expanded in powers of the Fermi momentum $k_f$ (modulo functions
of $k_f/m_\pi$). The empirical saturation point, the nuclear matter
incompressibility, and the asymmetry energy at saturation can be well
reproduced at order $\mathcal{O}(k_f^5)$ in the chiral expansion with just
one single momentum cut-off scale $\Lambda \simeq 0.65$ GeV 
which parameterizes short-distance physics.
In this approach the nuclear matter saturation mechanism is entirely
determined by in-medium pion-nucleon dynamics:
the interplay between attraction from two-pion exchange processes
and the stabilizing effects of the Pauli principle.
The implicit assumption made in the
analysis of Ref.~\cite{KFW.01} is the following:
there is no net contribution to the energy density
around saturation from the nucleon isoscalar self-energies, $\Sigma_S^{(0)}$
and $\Sigma_V^{(0)}$, that arise through changes in the quark condensate
and the quark density at finite baryon density.
At first sight, this assumption appears to be unrealistic,
given that in-medium QCD sum rules suggest individually
large scalar and vector self-energies of about 300 -- 400 MeV
in magnitude from the condensate background (see~Eqs.~(\ref{back1})-
(\ref{back2})).
On the other hand, the ratio
\begin{equation}
\label{ratio}
   \frac{\Sigma_S^{(0)}}{\Sigma_V^{(0)}} = - \frac{\sigma_N}{4(m_u +m_d)}
   \frac{\rho_s}{\rho}
\end{equation}
is indeed consistent with $-1$, for typical values of
the nucleon sigma term $\sigma_N$ and the current quark masses
$m_{u,d}$, and around nuclear matter saturation density 
where $\rho_s \simeq \rho$ (as an example, 
take $\sigma_N \simeq  50$ MeV~\cite{Gas} and $m_u +m_d \simeq 12$ MeV
\cite{Pic} at a scale of 1GeV).
One should note of course that the QCD sum rule constraints
implied by (\ref{back1})-(\ref{back2}) and by the ratio (\ref{ratio})
are not very accurate at a quantitative level.
The leading-order Ioffe formula on which Eq. (\ref{back1}) relies 
has corrections from condensates of higher dimension 
which are not well under control.
The estimated error in the ratio $\Sigma_S^{(0)}/\Sigma_V^{(0)}\simeq -1$ is 
about 20\%, given the uncertainties in the values of $\sigma_N$ and $m_u+m_d$.
Nevertheless, the constraints implied by Eq. (\ref{ratio}) give
important hints for further orientation.

%%%%%%%%%%%%%%%%%%%%%%%%%%%%%%%%%%%%%%%%%%%%%%%%%%%%%%%%%%%%%%%%%%%%%%%%%%%%

\subsection{\label{secIII-3}Equation of state 
based on in-medium Chiral Perturbation Theory}

%%%%%%%%%%%%%%%%%%%%%%%%%%%%%%%%%%%%%%%%%%%%%%%%%%%%%%%%%%%%%%%%%%%%%%%%%%%%

\begin{figure}
\includegraphics[scale=0.55,angle=0]{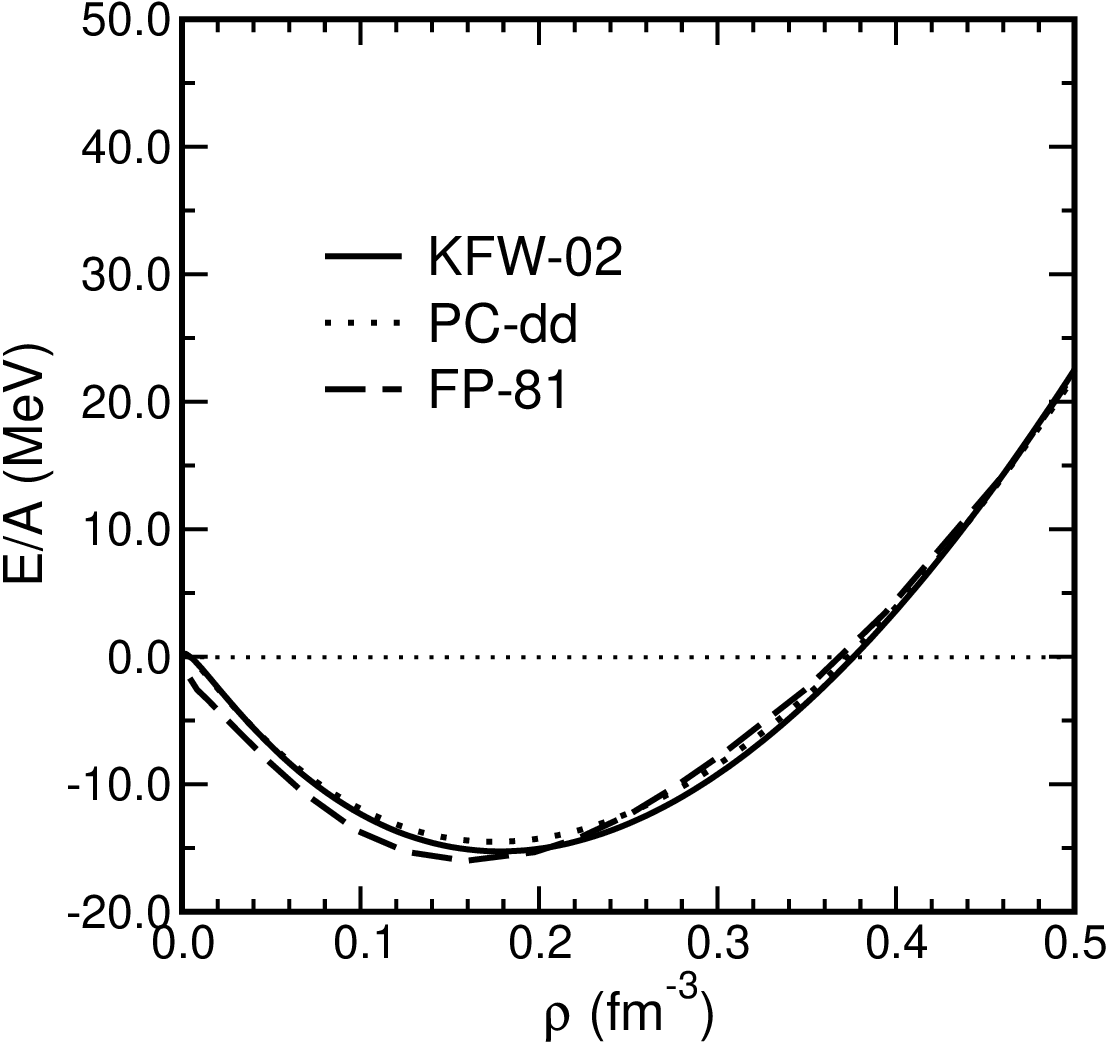}
\caption{\label{figA}Binding energy per nucleon for symmetric nuclear
matter as a function of baryon density. The solid curve (KFW-02)
is the EOS calculated in Ref.~\protect\cite{KFW.01} by using
in-medium CHPT for one- and two-pion exchange between nucleons.
The EOS displayed by the dotted curve (PC-dd) is obtained when the resulting
CHPT nucleon self-energies are mapped on the self-energies
of the relativistic point-coupling model with density dependent
couplings. FP-81 denotes the microscopic many-body
nuclear matter equation of state of
Friedman and Pandharipande~\protect\cite{FP.81}, shown for comparison.}
\end{figure}

In this work the nuclear matter equation of state will be calculated
in three steps. In the first approximation we follow the implicit assumption
made in Ref.~\cite{KFW.01} that $\Sigma_S^{(0)} = - \Sigma_V^{(0)}$
in nuclear matter, and neglect the contribution of the
condensate background self-energies to the nuclear matter 
equation of state (EOS).
The density dependence of the strength parameters is determined
by equating the isoscalar-scalar, the isoscalar-vector, the
isovector-scalar, and the isovector-vector
self-energies Eqs.~(\ref{self1}) -- (\ref{self8})
in the single-nucleon Dirac equation (\ref{Dirac}) with
those calculated using in-medium chiral perturbation theory:
\begin{eqnarray}
   G_S^{(\pi)} \rho_s & = & \Sigma_S^{\rm CHPT}(k_f,\rho) 
   \label{GS} \; ,\quad\quad\\
\label{49}
   G_V^{(\pi)} \rho + \Sigma_r^{(\pi)}
   & = & \Sigma_V^{\rm CHPT}(k_f,\rho) \label{GV} \; ,\\
   G_{TS}^{(\pi)} \rho_{s3} & = & 
   \Sigma_{TS}^{\rm CHPT}(k_f,\rho) \label{GTS} \; ,\\
   G_{TV}^{(\pi)} \rho_{3} & = & 
   \Sigma_{TV}^{\rm CHPT}(k_f,\rho) \label{GTV} \; ,
\end{eqnarray}
with
\begin{equation}
   \Sigma^{(\pi)}_r = \frac{1}{2} 
   \frac{\partial G_S^{(\pi)}}{\partial \rho}\rho_s^2
   + \frac{1}{2} \frac{\partial G_V^{(\pi)}}{\partial \rho} \rho^2 +
   \frac{1}{2} \frac{\partial G_{TS}^{(\pi)}}{\partial \rho}\rho_{s3}^2
   + \frac{1}{2} \frac{\partial G_{TV}^{(\pi)}}{\partial \rho} \rho_{3}^2 \; ,
\end{equation}
where the superscripts $\pi$ 
indicates that the density dependence of the couplings
originates from pionic (chiral) fluctuations.
The EOS of symmetric and asymmetric nuclear matter calculated in CHPT
give, via the Hugenholtz-van Hove theorem, the sums
$U_{(T)}(k_f,k_f) = \Sigma_{(T)S}^{\rm CHPT}(k_f,\rho)+
\Sigma_{(T)V}^{\rm CHPT}
(k_f,\rho)$
of the scalar and vector nucleon self-energies  
in the isoscalar and isovector channels
at the Fermi surface $p=k_f$ up to two-loop order, generated by
chiral one- and two-pion exchange \cite{KFW.02}.
The difference $\Sigma_{(T)S}^{\rm CHPT}(k_f,\rho)-
\Sigma_{(T)V}^{\rm CHPT} (k_f,\rho)$ is
calculated from the same pion-exchange diagrams via 
charge conjugation of the single particle potential in nuclear matter,
as explained in Ref.~\cite{KFW.02}.
Following a procedure similar to the
determination of the nucleon-meson vertices of relativistic mean-field models
from Dirac-Brueckner calculations \cite{HKL.01}, we neglect the momentum
dependence of $\Sigma_{(T)S,V}^{\rm CHPT}(p,\rho)$ and take their values at the
Fermi surface $p=k_f$. A polynomial fit up to order $k_f^5$ is performed,
and all four CHPT self-energies have the same functional form,
\begin{eqnarray}
   \Sigma(k_f, \lambda) & = & \left[ c_{30} + c_{31}\lambda +
   c_{32}\lambda^2 + c_{3L} \ln \frac{m_{\pi}}{4\pi f_{\pi}\lambda} \right]
   \frac{k_f^3}{M^2}\nonumber\\
& ~ & + c_{40} ~ \frac{k_f^4}{M^3} + \left[ c_{50} + c_{5L}
   \ln \frac{m_{\pi}}{4\pi f_{\pi}\lambda} \right] \frac{k_f^5}{M^4} \; ,
\label{expansion}
\end{eqnarray}
where the dimensionless parameter $\lambda$ is related to the
momentum cut-off scale by
$\Lambda = 2\pi f_{\pi} \lambda$,
and $f_\pi =$ 92.4 MeV denotes the pion decay constant.
In Table \ref{tab1} we display the coefficients of the expansion
(\ref{expansion}) of the CHPT self-energies for all four channels.
The cut-off parameter $\Lambda =$ 646 MeV that has been adjusted to the
nuclear matter saturation in Ref.~\cite{KFW.01}, corresponds to
$\lambda \simeq 1.113$. In order to determine the density dependence
of the couplings from Eqs. (\ref{GS}) -- (\ref{GTV}), the CHPT
self-energies are re-expressed in terms of the baryon
density $\rho = 2k_f^3/3\pi^2$:
\begin{eqnarray}
  \Sigma_S^{\rm CHPT}(k_f,\rho) & = & (c_{s1} + c_{s2}\rho^{\frac{1}{3}}
  + c_{s3}\rho^{\frac{2}{3}}) \rho \; ,\label{prho1} \\
  \Sigma_V^{\rm CHPT}(k_f,\rho) & = & (c_{v1} + c_{v2}\rho^{\frac{1}{3}}
  + c_{v3}\rho^{\frac{2}{3}}) \rho \; ,\label{prho2}\\
  \Sigma_{TS}^{\rm CHPT}(k_f,\rho) & = & (c_{ts1} + c_{ts2}\rho^{\frac{1}{3}}
  + c_{ts3}\rho^{\frac{2}{3}}) \rho_3 \; ,\label{prho3}\\
  \Sigma_{TV}^{\rm CHPT}(k_f,\rho) & = & (c_{tv1} + c_{tv2}\rho^{\frac{1}{3}}
  + c_{tv3}\rho^{\frac{2}{3}}) \rho_3 \label{prho4}\; .
\end{eqnarray}
The resulting expressions for the density dependent couplings
of the pionic fluctuation terms are\footnote{
small differences between $\rho_s$ and $\rho$ at nuclear matter densities
are neglected here
}
\begin{eqnarray}
  G_S^{(\pi)} & = & c_{s1} + c_{s2} \rho^{\frac{1}{3}}
  + c_{s3} \rho^{\frac{2}{3}} \; ,\\
\label{58}
  G_V^{(\pi)} & = & \bar{c}_{v1} + \bar{c}_{v2} \rho^{\frac{1}{3}}
  + \bar{c}_{v3} \rho^{\frac{2}{3}}
  \quad \left\{ \begin{array}{ccl}
\bar{c}_{v1} & = & c_{v1} \\
\bar{c}_{v2} & = & \frac{1}{7}(6c_{v2}-c_{s2} - \delta^2(c_{ts2} +c_{tv2}) )\\
\bar{c}_{v3} & = & \frac{1}{4}(3c_{v3}-c_{s3} - \delta^2(c_{ts3} +c_{tv3}))
\end{array} \right. \; ,\\
  G_{TS}^{(\pi)} & = & c_{ts1} + c_{ts2} \rho^{\frac{1}{3}}
  + c_{ts3} \rho^{\frac{2}{3}} \; ,\\
  G_{TV}^{(\pi)} & = & c_{tv1} + c_{tv2} \rho^{\frac{1}{3}}
  + c_{tv3} \rho^{\frac{2}{3}} \; ,
\end{eqnarray}
where $\delta = (\rho_n - \rho_p)/(\rho_n + \rho_p)$.
For $\lambda \simeq 1.113$, the coefficients
of the expansion of the CHPT self-energies in powers
of the baryon density Eqs. (\ref{prho1}) -- (\ref{prho4}) are given
in Table \ref{tab2}. In Fig.~\ref{figA} the resulting equation of state
of isospin symmetric nuclear matter, calculated from
Eqs. (\ref{epsilon}) and (\ref{pressure}), is compared
with the CHPT nuclear matter EOS of Ref.~\cite{KFW.01}.
In Table \ref{tab3} we compare the corresponding
nuclear matter properties at the saturation point:
the binding energy per particle, the saturation density, the
compression modulus, and the asymmetry energy.
The point-coupling model, with the density dependence of the strength
parameters of the second order interaction terms determined
by pionic (chiral) fluctuations, nicely reproduces the
nuclear matter EOS calculated using in-medium CHPT. Small
differences arise mainly because in the mapping of the
CHPT nucleon self-energies on the self-energies of the
point coupling model in Eqs. (\ref{GS}) -- (\ref{GTV}), the
momentum dependence of the former has been frozen to the
values at the Fermi momentum. This is a well known problem that arises
also in the determination of the meson-nucleon in-medium vertices
from Dirac-Brueckner calculations of nucleon self-energies
in nuclear matter~\cite{HKL.01}.
For comparison in Fig.~\ref{figA} we also include the microscopic many-body
nuclear matter equation of state of
Friedman and Pandharipande~\cite{FP.81}.
\par
\begin{figure}
\includegraphics[scale=0.55,angle=0]{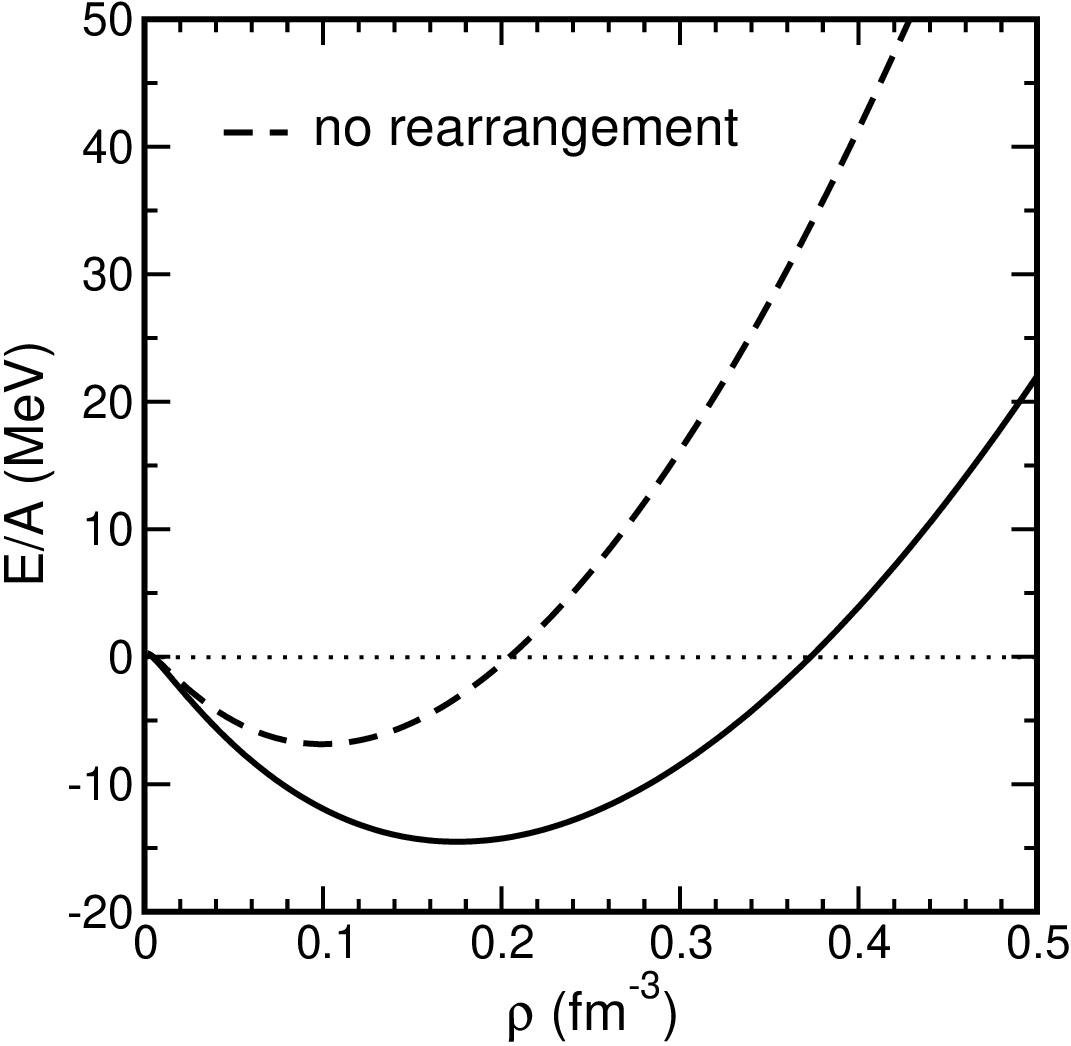}
\caption{\label{figB} Equation of state of symmetric nuclear matter
calculated in the relativistic point-coupling model with and
without the rearrangement contribution to the isoscalar
vector self-energy.}
\end{figure}

The importance of the rearrangement self-energies
implemented through Eqs. (\ref{49}) and (\ref{58}) for a
model with density dependent couplings is illustrated in Fig.~\ref{figB},
where we compare the EOS of isospin symmetric nuclear matter calculated
with and without the rearrangement contributions in the Eq.~(\ref{GV})
that determines the density dependence of the vertex of the
isoscalar-vector interaction term. The difference is indeed very large.
Without the rearrangement terms in Eq.~(\ref{GV})
a saturation density of only 0.1 fm$^{-3}$ is obtained, and the
binding energy per particle is $-6.85$ MeV. Also the nuclear matter
incompressibility is too low: 124 MeV. Obviously,
only with inclusion of the rearrangement contributions in the
isoscalar-vector self-energy it is possible to reproduce the
nuclear matter EOS calculated in CHPT in an equivalent approach
with density dependent point couplings.

At very low-densities, the CHPT equation of state 
(or any mean-field approach) cannot reproduce
realistic calculations of $E/A$, for obvious reasons.
The low density limit of the energy per particle is not given by a free
Fermi gas of nucleons, but by a gas of clusters (deuterons, etc$\ldots$).
As a consequence, the low-density limit of $E/A$ approaches a constant
as $k_f \rightarrow 0$.
The effect of this shift is marginal, however, for the bulk
of nuclear matter around saturation density.

%%%%%%%%%%%%%%%%%%%%%%%%%%%%%%%%%%%%%%%%%%%%%%%%%%%%%%%%%%%%%%%%%%%%

\subsection{Constraints from QCD Sum Rules: leading orders}

%%%%%%%%%%%%%%%%%%%%%%%%%%%%%%%%%%%%%%%%%%%%%%%%%%%%%%%%%%%%%%%%%%%%

The next step in our calculation of the nuclear matter EOS  
includes the contributions of the condensate background self-energies
in the isoscalar-scalar and isoscalar-vector channels.
In Ref.~\cite{Finelli:2002na} we have shown that, even though in a
first approximation the condensate potentials do not play a role
in the saturation mechanism, they are indeed essential for the
description of ground-state properties of finite nuclei.

The nuclear matter EOS is calculated from  
Eqs. (\ref{epsilon}) and (\ref{pressure}). In this case,
the strength parameters of the isoscalar-scalar and isoscalar-vector
interaction terms have the form given in Eq.~(\ref{couplings}), i.e.
they include contributions from condensate background fields.
For the isovector channel, on the other hand,
the assumption of the model is that only pionic (chiral)
fluctuations contribute to the nucleon self-energies.
In leading order of the in-medium QCD sum rules the nucleon
isoscalar self-energies that arise through changes in the
quark condensate and the quark density are linear functions of the
corresponding scalar and baryon densities
(see Eqs. (\ref{sigmaS}) and (\ref{sigmaV})).
To this order, therefore, the corresponding couplings $G_{S,V}^{(0)}$ should
be density independent.
Eq. (\ref{back1}) implies (identifying $M$ with the free nucleon mass):
\begin{equation}
\Sigma_S^{(0)} \simeq -350~{\rm MeV}~\frac{\sigma_N}
{50~{\rm MeV}} \,{\rho_s\over \rho_0} \; .
\end{equation}
Evidently, the leading-order in-medium change of the chiral condensate
is a source of a strong, attractive scalar field which acts
on the nucleon in such a way as to reduce its mass in nuclear matter
by more than $1/3$ of its vacuum value.
Using (\ref{sigmaS}) one estimates:
\begin{equation}
\label{estimate}
G_S^{(0)} \simeq  -11~{\rm fm}^2~\frac{\sigma_N}{50~{\rm MeV}} \quad {\rm at}
\,\,\,\rho_s \simeq \rho_0 = 0.16~{\rm fm}^{-3}.
\end{equation}
A correspondingly smaller value of $G_{S}^{(0)}$ results if
only a fraction of the nucleon mass $M$ is associated with the chiral
condensate, leaving room for non-leading
contributions from higher dimensional condensates.
Since the QCD sum rules suggest that the ratio of the condensate scalar and
vector self-energies is close to $-1$, one expects roughly
$G_V^{(0)} \simeq - G_S^{(0)}$.
\par
In our model both $G_S^{(0)}$ and
$G_V^{(0)}$ are parameters to be adjusted to ``empirical" saturation
properties of nuclear matter, as well as to ground-state properties
of finite nuclei. Moreover, since the nuclear matter saturation mechanism
is not any more exclusively determined by pionic (chiral) fluctuations,  
the momentum cut-off scale $\Lambda$ (or the dimensionless $\lambda$)
is an additional parameter in our calculation of the nuclear matter EOS.
At this preliminary stage, the three parameters: $G_S^{(0)}$, $G_V^{(0)}$ 
and $\Lambda$ are adjusted to reproduce ``empirical" nuclear matter properties:
$E/A = -16$ MeV (5\%), $\rho_0 = 0.153$ fm$^{-3}$  (10\%),
$K_0 = 250$ MeV (10\%), and $a_4 = 33$ MeV (10\%). The values in  
parentheses correspond to the error bars used in the fitting procedure.

\begin{figure}
\includegraphics[scale=0.55,angle=0]{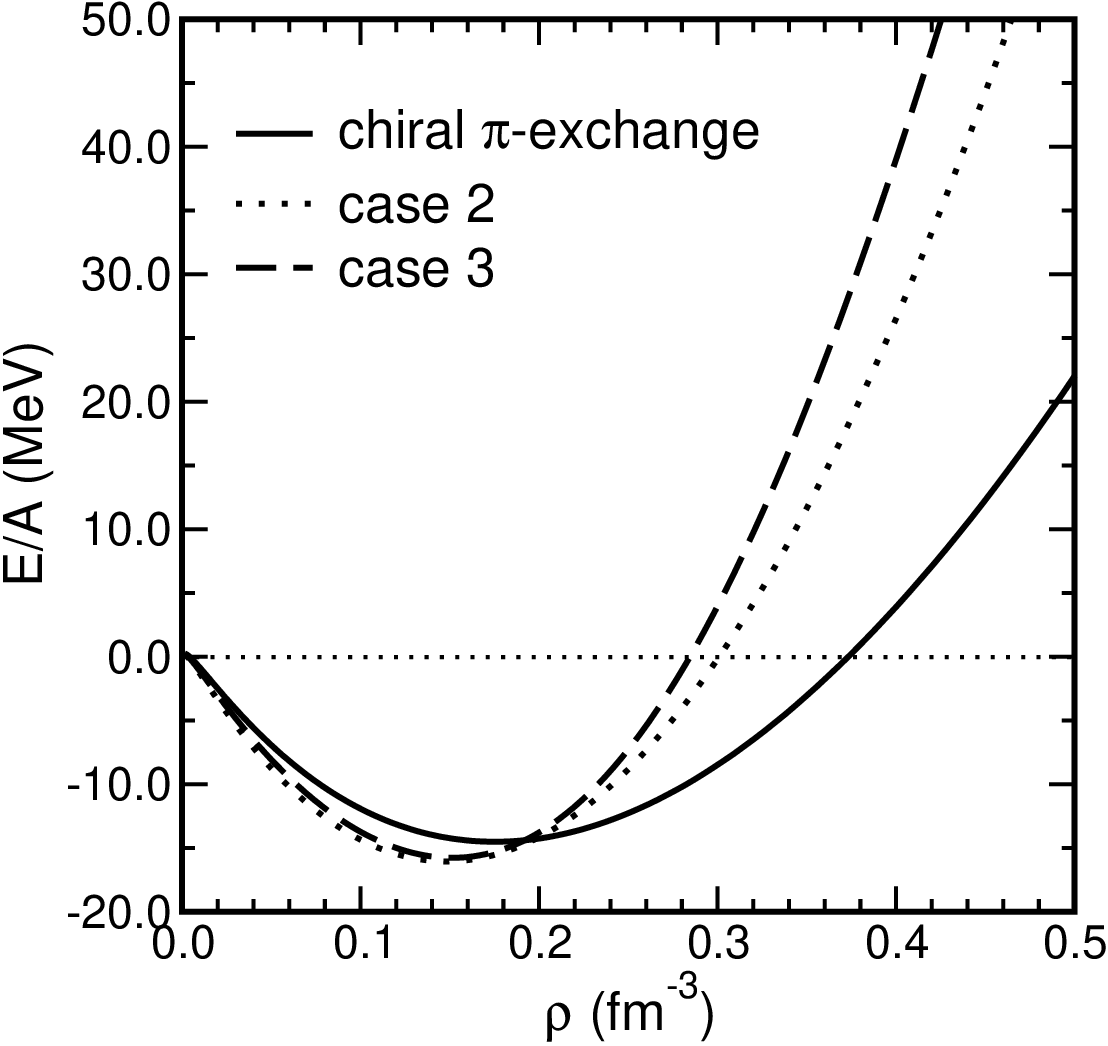}
\caption{\label{figC} Binding energy per nucleon for symmetric nuclear
matter as a function of the baryon density, calculated from
chiral one- and two-pion exchange between nucleons (case 1, solid curve),
by adding the isoscalar condensate background nucleon self-energies 
linear in the
corresponding densities, with $G_V^{(0)} = - G_S^{(0)} = 7~{\rm fm}^2$
(case 2, dotted curve), and finally by including also the non-linear 
contribution
$g_V^{(1)} \rho^2$ to the isoscalar vector condensate self-energy,
with $G_V^{(0)} = 11~{\rm fm}^2$, $G_S^{(0)} = -12~{\rm fm}^2$
(case 3, dashed curve).}
\end{figure}

The result of the corresponding {\it least-squares fit} is shown in
Fig.~\ref{figC} (EOS of isospin symmetric nuclear matter in
comparison with the one calculated exclusively from chiral pion-exchange)
and in the first row of Table \ref{tab4} for $G_S^{(0)} =  -7$ fm$^2$,
$G_V^{(0)} =  7$ fm$^2$, and $\Lambda = 685$ MeV. The EOS so obtained
is not yet satisfactory. The asymmetry energy at saturation is
too high, and the values of $G_S^{(0)}$ and  $G_V^{(0)}$
are smaller than the leading order QCD sum rule
estimate Eq. (\ref{estimate}). This results in the relatively large Dirac
effective mass $M^*/M =$ 0.75. A high Dirac mass in turn indicates
that the effective single-nucleon spin-orbit potential is too
weak to reproduce the empirical energy spacings between
spin-orbit partner states in finite nuclei.
Stronger background scalar and vector fields are required in order  
to drive the large spin-orbit splittings in finite nuclei.

Notice however that, even though
$G_S^{(0)}$ and $G_V^{(0)}$ have been varied independently, the
minimization procedure prefers to balance the contributions from the
corresponding large scalar and vector self-energies so that their
sum tends to vanish.
There is already enough binding from pionic fluctuations alone,
and therefore $\Sigma_S^{(0)} = - \Sigma_V^{(0)}$ represents a
very good approximation for the condensate potentials.
This remarkable feature prevails as we now proceed
with fine-tuning improvements.

%%%%%%%%%%%%%%%%%%%%%%%%%%%%%%%%%%%%%%%%%%%%%%%%%%%%%%%%%%%%%%%%%%%%%%%

\subsection{Corrections of higher order}

%%%%%%%%%%%%%%%%%%%%%%%%%%%%%%%%%%%%%%%%%%%%%%%%%%%%%%%%%%%%%%%%%%%%%%%

Up to this point the nucleon self-energies are evaluated
at two-loop level including terms of order $k_f^5$. At order
$k_f^6$ ($\propto \rho^2$), several additional effects appear.
First, condensates of higher dimension enter, such as
$\langle\bar{q}\Gamma q \bar{q} \Gamma q \rangle$,
$\langle q^\dagger q\rangle^2$ and the gluon condensate, the detailed
in-medium dependence of which is difficult to estimate.
Secondly, four-loop CHPT contributions to the energy density
(three-loop in the self energies) introduce genuine
3-body interactions.
All those effects combine to produce the $\mathcal{O} (k_f^6)$ correction
in the energy per particle.
Our chiral counting conjecture is that these terms should be subleading in
the $k_f$-expansion.
In estimating those contributions, we will take
a pragmatic point of view.
Given the density-independent leading condensate terms $G^{(0)}$
and the CHPT couplings $G^{(\pi)}(\rho)$ evaluated to
$\mathcal{O} (k_f^5) \propto \rho^{\frac{2}{3}}$, we generalize
\begin{equation}
G(\rho) = G^{(0)} + G^{(\pi)} (\rho) + \delta G^{(1)} (\rho) \; ,
\end{equation}
adding terms $\delta G^{(1)} = g^{(1)} \rho$ and let the constant $g^{(1)}$ be
determined by a least squares fit to properties of nuclear matter and
finite nuclei. At first sight, the corrections $\delta G^{(1)}_S,~
\delta G^{(1)}_V,{\rm etc}\ldots$ would then introduce 
several additional parameters, undermining our startegy to keep 
the number of freely adjustable fine-tuning constants as small as possible. 
For the scalar self-energy a primary uncertainty arises from the 
four-quark condensate
and its assumed factorization into the form $\langle \bar{q} q \rangle^2$,
as discussed in detail in Ref.~\cite{Jin.94}. While the in-medium values
of condensates such as $\langle\bar{q}\Gamma q \bar{q} \Gamma q \rangle$
are not well determined, it nevertheless appears that only a weak
or moderate density dependence is consistent with known nuclear phenomenology.
We do not have to be much concerned about this issue since our approach
explicitly includes scalar $\pi-\pi$ fluctuations
(up to order $k_f^5$ in the energy density), which should account for
a large part of the four quark condensate effects.
Remaining short-distance effects can be absorbed in $\delta G^{(1)}$.

As a first option, assume that the only non-vanishing $\delta G^{(1)}$
is $\delta G^{(1)}_V = g^{(1)}_V \rho$, and determine the
constant $g^{(1)}_V$.
For $G_S^{(0)} = -12$ fm$^2$,
$G_{V}^{(0)} =  11$ fm$^2$, $g_{V}^{(1)} =  -3.9$ fm$^5$ and
$\Lambda = 600$ MeV, determined in a {\it least-squares fit}
to the "empirical" nuclear matter input,
the EOS resulting is displayed in Fig.~\ref{figC}
and the nuclear matter properties at saturation are listed
in the second row of Table \ref{tab4}. This nuclear matter EOS
is actually quite satisfactory, with realistic values of the
binding energy, saturation density
and a low effective Dirac mass. The only exception
is a relatively large nuclear matter incompressibility.
Calculations of the excitation energies of
isoscalar giant monopole resonances in
spherical nuclei in the time-dependent relativistic mean-field
framework~\cite{Vre.97}, and in the relativistic random-phase
approximation~\cite{Ma.01}, suggest that the nuclear matter
incompressibility modulus should be in the
range $K_0 \approx 250 - 270$ MeV. A value of $K_0$ in this
range, however, can only be obtained by including the
next higher order non-linear self-interaction term
in the expansion of $\Sigma_V$ or $\Sigma_S$,
i.e. a term proportional to $\rho^3$. This feature is also
well known in standard meson-exchange and point-coupling
relativistic mean-field models. In the present
analysis we do not aim for a detailed description of
nuclear properties that depend crucially
on the incompressibility, and we also wish to keep the number
of adjustable parameters at minimum, so the density dependence of
$G_V$ is parameterized as
\begin{equation}
  G_V (\rho) = G_{V}^{(0)} + G_{V}^{(\pi)} (\rho) + g_V^{(1)} \rho \; .
\end{equation}
The resulting couplings of the condensate background
fields ($G_S^{(0)} = - 12$ fm$^2$ and $G_{V}^{(0)} =  11$ fm$^2$)
are remarkably close to the prediction (\ref{estimate}) of the
leading order in-medium QCD sum rules.
The scalar and the vector ``condensate''
self-energies $\Sigma_S^{(0)}=G_S^{(0)}\rho_s$
and $\Sigma_V^{(0)}=G_V^{(0)}\rho$ follow the expectation
$\Sigma_S^{(0)} \simeq -\Sigma_V^{(0)}$ to within
5\% at saturation density, even without this condition
being pre-imposed.  
The large isoscalar condensate background self-energies
in turn lead to a relatively low
effective Dirac mass, crucial for the empirical spin-orbit
splittings in finite nuclei. Finally, the cut-off  
$\Lambda = 600$ MeV differs by less than 10 \% from the value
(646 MeV) obtained when the nuclear matter EOS results
exclusively from one- and two-pion exchange between
nucleons~\cite{KFW.01}.
\par
The $\delta G^{(1)}_V = g^{(1)}_V\rho$ term acts like a three-body force
in the energy density. Its effect is relatively small: at saturation density,
$\delta G^{(1)}_V/G^{(0)}_V \simeq 0.05$.
Splitting $\delta G^{(1)}$ between scalar and vector parts
(e.g. by choosing $\delta G^{(1)}_V \simeq \delta G^{(1)}_S
\simeq -2~{\rm fm}^{-5}$) would induce only marginal changes.

%%%%%%%%%%%%%%%%%%%%%%%%%%%%%%%%%%%%%%%%%%%%%%%%%%%%%%%%%%%%%%%%%%%%%%%

\subsection{Asymmetric nuclear matter}

%%%%%%%%%%%%%%%%%%%%%%%%%%%%%%%%%%%%%%%%%%%%%%%%%%%%%%%%%%%%%%%%%%%%%%%

The energy per particle of asymmetric nuclear
matter can be expanded about the equilibrium density $\rho_{\rm sat}$
in a Taylor series in $\rho$ and $\delta$~\cite{Lee.98}
\begin{equation}
   E(\rho,\delta) = E(\rho,0) + S_2(\rho) \delta^2 
   + S_4(\rho) \delta^4 + \cdots
\label{taylor}
\end{equation}
where $\delta = (\rho_n - \rho_p)/(\rho_n + \rho_p)$, and 
\begin{eqnarray}
   E(\rho,0) ~& = & - a_v + \frac{K_0}{18 \rho_{\rm sat}^2}~
   (\rho - \rho_{\rm sat})^2 + ... \; ,\\
   S_2(\rho) ~& = &~ a_4 + 
   \frac{p_0}{\rho_{\rm sat}^2}~(\rho - \rho_{\rm sat}) +
   \frac{\Delta K_0}{18 \rho_{\rm sat}^2}~ (\rho - \rho_{\rm sat})^2 
   + \cdots \; .
\label{S2}
\end{eqnarray}

The empirical value at saturation density is $S_2(\rho_{\rm sat}) = a_4 =
30\pm 4$ MeV. The parameter $p_0$ defines the linear density dependence
of the symmetry energy, and $\Delta K_0$ is the correction to the
incompressibility. The contribution of the term $S_4(\rho) \alpha^4$
in (\ref{taylor}) is very small
in ordinary nuclei and it will be neglected in the present work.
In a recent analysis of neutron
radii in non-relativistic and covariant mean-field models~\cite{Fur.01},
Furnstahl has studied the linear correlation between the neutron skin
and the asymmetry energy. In particular, he has shown that there is a
very strong linear correlation between the neutron skin thickness in
$^{208}$Pb and the individual parameters that determine the
asymmetry energy $S_2(\rho)$: $a_4$, $p_0$ and $\Delta K_0$. The empirical
value of the difference between neutron and proton {\it rms} radii
$r_n - r_p$ in $^{208}$Pb ($0.20 \pm 0.04$ fm from proton
scattering data~\cite{SH.94}, and $0.19 \pm 0.09$ fm from the alpha scattering
excitation of the isovector giant dipole resonance~\cite{Kra.94}) places the
following constraints on the values of the parameters of the symmetry
energy: $a_4 \approx 30-34$ MeV, 2 MeV/fm$^3 \leq p_0 \leq$ 4 MeV/fm$^3$,
and $-200$ MeV $\leq \Delta K_0 \leq -50$ MeV.

\begin{figure}
\includegraphics[scale=0.55,angle=0]{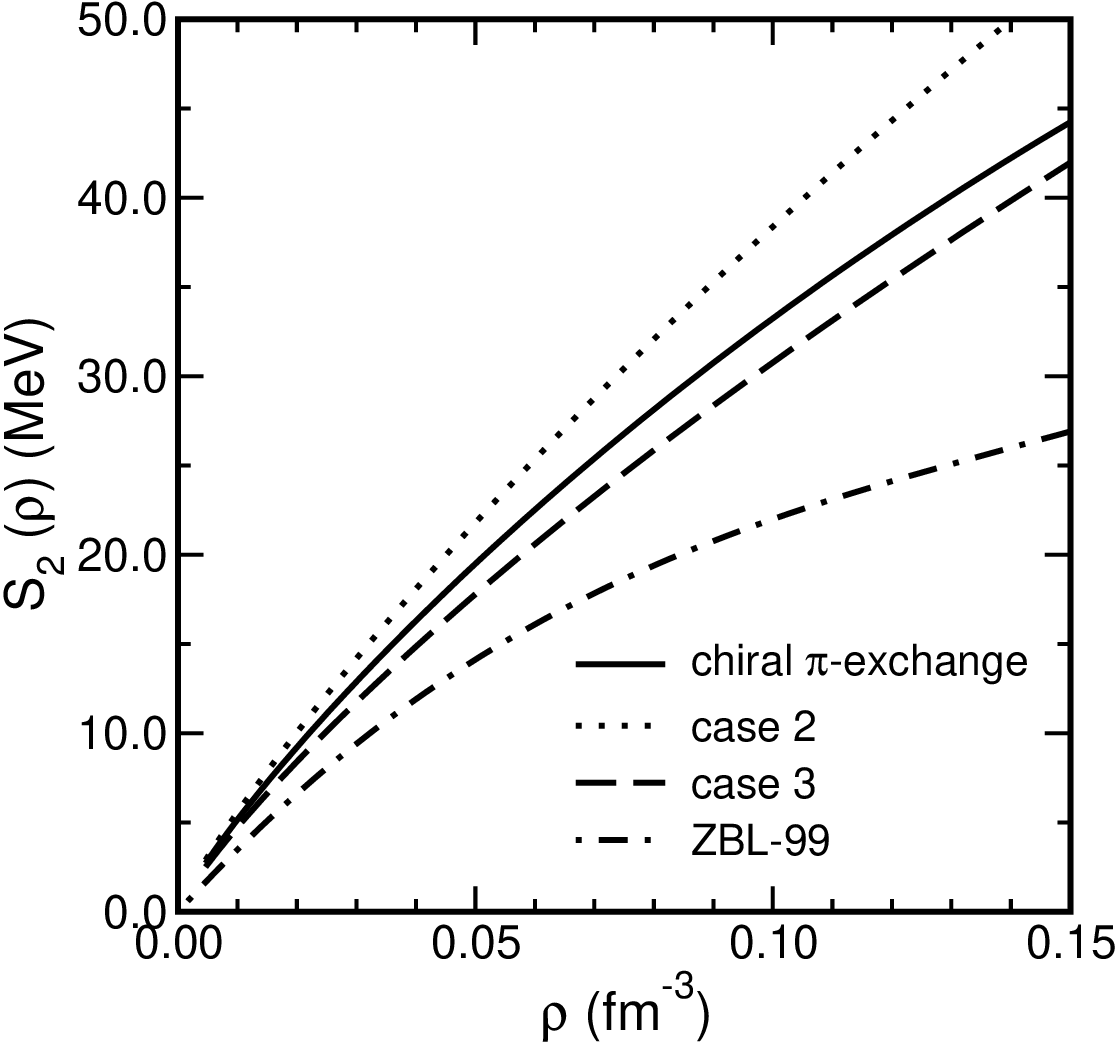}
\caption{\label{figD} $S_2(\rho)$ coefficient (\protect\ref{S2})
of the quadratic term of the energy per particle of asymmetric
nuclear matter. The three curves correspond to calculations with
only chiral one- and two-pion exchange between nucleons,
with the inclusion of leading order  
isoscalar condensate background nucleon self-energies,
and finally with the contribution of the non-linear
vector condensate nucleon self-energy.
Cases 2 and 3 are explained in the caption of Fig.~\ref{figC} and
ZBL-99 denotes the asymmetry energy curve
obtained in the non-relativistic Brueckner-Hartree-Fock calculation
of Ref.~\protect\cite{ZBL.99}.}
\end{figure}

In Fig.~\ref{figD} we plot the asymmetry energy $S_2(\rho)$ as function
of the baryon density for the three steps of our calculation of the
nuclear matter EOS: first, using only chiral one- 
and two-pion exchange calculated in CHPT, next with the 
inclusion of leading order isoscalar condensate background 
nucleon self-energies that arise through
changes in the quark condensate and the quark density
at finite baryon density, and finally with the non-linear
contribution $\delta G_V^{(1)}$ to the vector condensate nucleon self-energy. 
The three curves can be compared, for instance, with the asymmetry energy
obtained in the non-relativistic Brueckner-Hartree-Fock calculation
of Ref.~\cite{ZBL.99}, or with the asymmetry energy of the relativistic
meson-exchange model with a phenomenological density dependence of the
meson-nucleon couplings in Ref.~\cite{Niksic:2002yp}. 

\begin{figure}
\includegraphics[scale=0.55,angle=0]{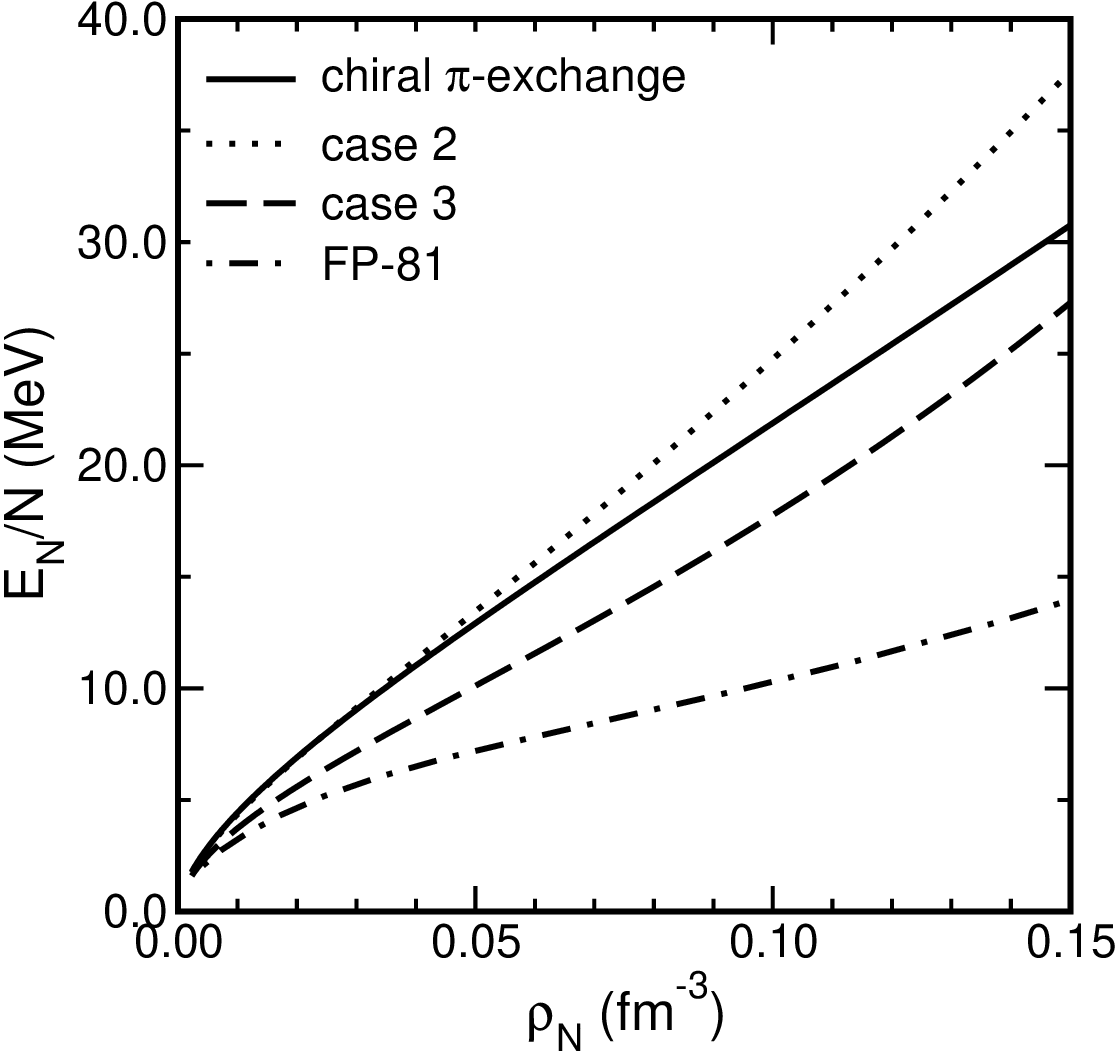}
\caption{\label{figE} Energy per particle of neutron matter as a
function of the neutron density. The results are obtained from
chiral one- and two-pion exchange between nucleons (solid curve),
by adding the isoscalar condensate background nucleon self-energies linear
in the corresponding densities (case 2, dotted curve), and
finally by including also the non-linear contribution to the
isoscalar vector condensate self-energy (case 3, dashed curve).
Cases 2 and 3 are the same as in the caption of Fig.~\ref{figC}.
This is shown in comparison with the microscopic many-body
neutron matter equation of state of
Friedman and Pandharipande~\protect\cite{FP.81}.}
\end{figure}

In Fig.~\ref{figE} we display the energy per particle of neutron matter as
a function of the neutron density. The results obtained from
chiral one- and two-pion exchange between nucleons, by adding the
isoscalar condensate background nucleon self-energies linear in the
corresponding densities, and
finally by including the non-linear ``3-body'' contribution to the
isoscalar vector self-energy, are
shown in comparison with the microscopic many-body
neutron matter equation of state of
Friedman and Pandharipande~\cite{FP.81}. 

These comparisons show a qualitatively correct trend in both
the asymmetry energy and the neutron matter equation of state,
but refinements of the density dependent isovector couplings are
obviously still necessary.

Of course, one cannot expect to reproduce
the behaviour at very low neutron density in a perturbative approach.
The very large neutron-neutron scattering length requires
a partial resummation which effectively
renormalizes the slope of the $k_f^2$-term in $E_N/N$ by roughly a factor of
one-half~\cite{Fur.01,Ste.00}.
 
%%%%%%%%%%%%%%%%%%%%%%%%%%%%%%%%%%%%%%%%%%%%%%%%%%%%%%%%%%%%%%%%%%%%%%%

\subsection{Intermediate summary and discussions}

%%%%%%%%%%%%%%%%%%%%%%%%%%%%%%%%%%%%%%%%%%%%%%%%%%%%%%%%%%%%%%%%%%%%%%%

While the quantitative details for the terms contributing to the scalar
and vector self-energies can be reconstructed from the previous
sections using Tables \ref{tab1} and \ref{tab2}, it is instructive
to observe systematic trends in these self-energies when expanded
in powers of $k_f$ (or equivalently, in powers of $\rho^{\frac{1}{3}}$).
We focus on symmetric nuclear matter and write
\begin{equation}
   \Sigma_{S,V} = \Sigma_{S,V}^{(0)} + \Sigma_{S,V}^{(\pi)}
   +\delta\Sigma_{S,V} \; ,
\end{equation}
combining the leading condensate terms $\Sigma_{S,V}^{(0)}$,
the chiral (pionic) terms $\Sigma_{S,V}^{(\pi)}$ calculated
to order $k_f^5$, and possible corrections of higher order
summarized in $\delta \Sigma_{S,V}$.
We introduce the saturation density $\rho_0 = 0.16~{\rm fm}^{-3}$
as a convenient scale and write the condensate terms as
\begin{eqnarray}
   \Sigma_S^{(0)} (\rho) & \simeq & - 0.35~{\rm GeV}~ 
   \frac{|G_S^{(0)}|}{11~{\rm fm}^2}
   \left( \frac{\rho_s}{\rho_0} \right) \; ,\\
   \Sigma_V^{(0)} (\rho) & \simeq & + 0.35~{\rm GeV}~ 
   \frac{|G_V^{(0)}|}{11~{\rm fm}^2}
\left( \frac{\rho}{\rho_0} \right) \; .
\end{eqnarray}
We note that  with $|G_S^{(0)}|$ chosen less than 5\% larger than $G_V^{(0)}$,
this just compensates for the difference between the baryon density $\rho$
and the scalar density $\rho_s < \rho$ such that $\Sigma_S^{(0)}
\simeq - \Sigma_V^{(0)}$ results at $\rho = \rho_0$.
The ``best-fit'' values $G_S^{(0)} \simeq -12~{\rm fm}^2$ and
 $G_V^{(0)} \simeq + 11~{\rm fm}^2$ are thus perfectly consistent
with QCD sum rule expectations: a non-trivial result.
The pionic fluctuations from one- and two-pion exchange processes
have the following approximate pattern:
\begin{equation}
   \Sigma_{S,V}^{(\pi)} (\rho) \simeq - 
   75~{\rm MeV}~\left(\frac{\rho}{\rho_0} \right)
   \left[ 1 + d_1 \left(\frac{\rho}{\rho_0} \right)^{\frac{1}{3}} +
   d_2 \left(\frac{\rho}{\rho_0} \right)^{\frac{2}{3}}\right] \; ,
\end{equation}
where $d_1$ varies between $-0.61$ (scalar) and $-0.65$ (vector), and
$d_2 \simeq -0.17$ for both cases. Finally, the higher order 
corrections are summarized as
\begin{equation}
   \delta \Sigma_S + \delta \Sigma_V \simeq 
   -20~{\rm MeV}~\left( \frac{\rho}{\rho_0}
   \right)^2 \; ,
\end{equation}
where the detailed decomposition into scalar and vector part turns out not to
be relevant.

The leading attraction at $\mathcal{O}(k_f^3)$ in $\Sigma_{S,V}^{(\pi)}$
depends on the cut-off scale $\Lambda \equiv 2 \pi f_\pi \lambda$
which distinguishes ``active'' two-pion exchange dynamics 
at long and intermediate distances from
short-distance (high momentum) dynamics. Details at short distance
scales (related to the intrinsic structure of the nucleons and their
short-range interactions) are not resolved as long as the Fermi momentum
$k_f$ is much smaller than $4\pi f_\pi$. 
The unresolved short-distance information
can then be translated into an equivalent (density-independent)
four-point vertex which generates self-energy terms linear
in the density at the mean-field level.

\subsection{Comparison with the Dirac-Brueckner G matrix}

\begin{figure}
\includegraphics[scale=0.7,angle=0]{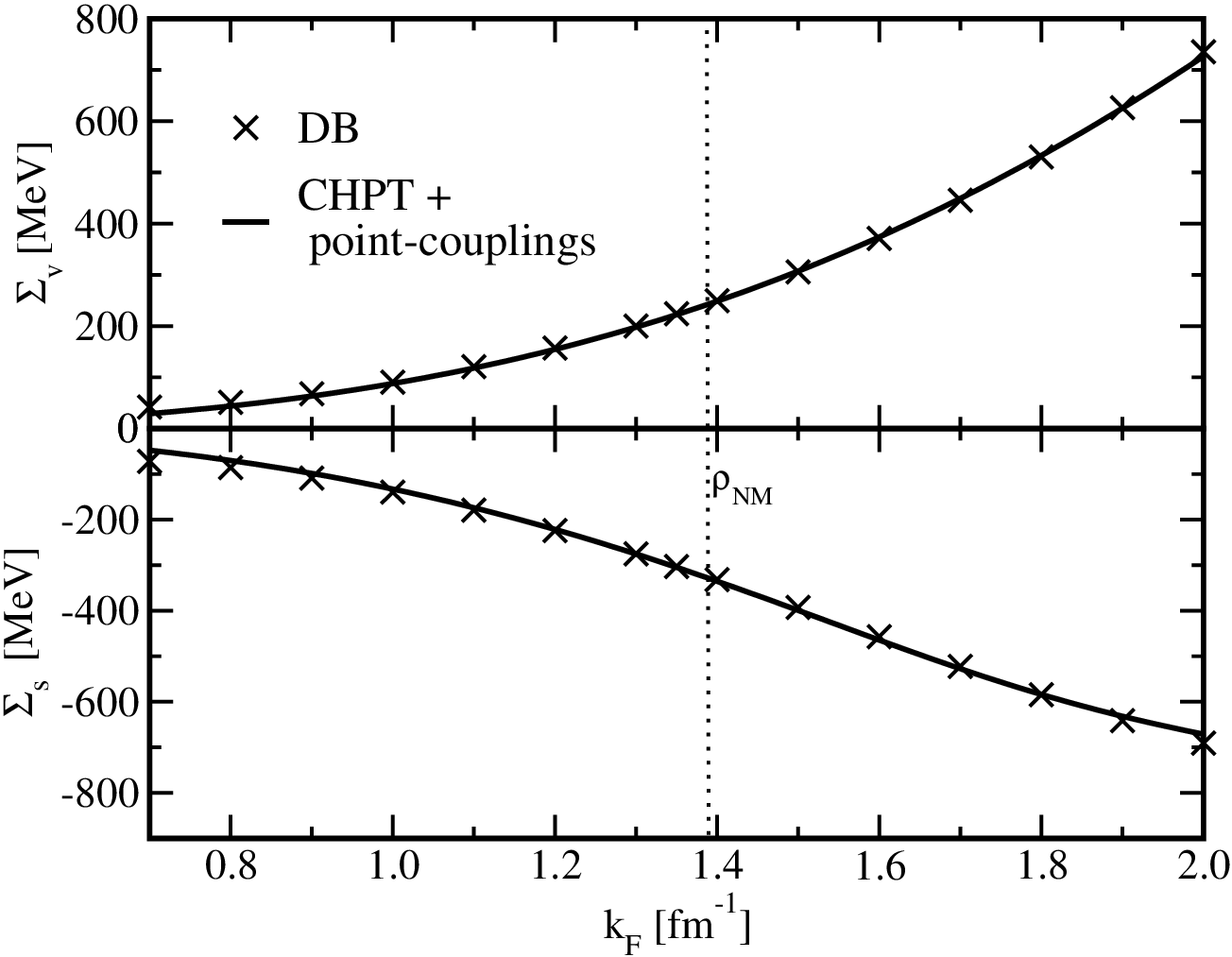}
\caption{\label{fig_DB}
Comparison of the $k_f$-dependence of the isoscalar vector
and scalar self-energies resulting from a Dirac-Brueckner
G-matrix calculation (solid lines: DB~\cite{Gro.98}), with the self-energies
generated from in-medium chiral perturbation theory
(dashed lines: CHPT + point-couplings up to 3-loop order in
the energy density). 
}
\end{figure}

At this point it is instructive to compare
the density-dependence of our self-energies $\Sigma_{S,V} (\rho)$
with that of the scalar and the vector self-energies resulting from full
Dirac-Brueckner calculations based on realistic NN-interactions.
We refer here explicitly to a calculation, reported in Ref.~\cite{Gro.98},
which uses the Bonn A potential. With this potential,
the nuclear matter saturation density comes out somewhat higher ($\rho_{sat}
\simeq 0.185~{\rm fm}^{-3}$) than the one of our best fit. Simulating this
higher saturation density in our approach requires weakening the
condensate mean fields $\Sigma_{S,V}^{(0)}$ by about $25$ \%, 
but with no changes in the pionic terms $\Sigma_{S,V}^{(\pi)}$. 
After this readjustment, the difference in the $k_f$-dependences 
between our $\Sigma_S$ and $\Sigma_V$ and those resulting from 
the Dirac-Bruckner calculation
is less than $10$ \% over the entire range of densities from $0.5\rho_0$
to $2.5\rho_0$~(see Fig.~\ref{fig_DB}).

This is a remarkable observation:
it appears that in-medium chiral perturbation theory at two-loop order,
with a cut-off scale $\Lambda \simeq 0.6$ GeV that converts
unresolved short-distance dynamics at momenta beyond $\Lambda$ effectively
into contact terms, generates quantitatively similar 
in-medium nucleon self energies
as a full Dirac-Brueckner calculation, when requiring that both approaches
reproduce the same nuclear matter saturation point. The reasoning behind this
observation can presumably be traced back to the separation of scales at work,
a key element of chiral effective field theory  on which our approach is based.
One- and two-pion exchange in-medium dynamics at momentum scales
comparable to $k_f$ are treated explicitly.
Our in-medium CHPT approach includes all terms (ladders and others)
to three-loop order in the energy density, a prominent one being
iterated one-pion exchange.
The non-trivial $k_f$
dependence in the self-energies (beyond ``trivial'' order $k_f^3$)
reflects the action of the Pauli principle in these processes. 
Such features are also present in leading orders of the Brueckner
ladder summation which includes, for example, Pauli blocking effects on 
iterated one-pion exchange.
These terms produce the characteristic $k_f^4$-dependence of the energy
per particle.
On the other hand, the iteration of short-distance interactions to 
all orders in the Brueckner ladder,
involves intermediate momenta much larger than $k_f$ and generates
similar self-energy pieces (proportional to density $\rho$) 
as a contact interaction reflecting the high-momentum scale $\Lambda$. 
Note that this contact interaction must {\it not} be iterated
as it already represents the full short-distance T-matrix information.
\par
The reasoning here has a close
correspondence to related studies in Ref. \cite{Bog.03}.

%=========================================================================
%  Section 4

\section{\label{secIV}Finite nuclei}
%=========================================================================

In this section the relativistic nuclear point-coupling model,
constrained by in-medium QCD sum rules and chiral perturbation theory,
is applied in calculations of ground state properties of finite nuclei.
In the mean-field approximation
the ground state of a nucleus with A nucleons is represented by
the antisymmetrized product of the lowest occupied single-nucleon
stationary solutions of the Dirac equation~(\ref{Dirac}).
The calculation is self-consistent in the sense that the
nucleon self-energies are functions of nucleon densities and currents,
calculated from the solutions of Eq.~(\ref{Dirac}).
The mean-field approach to nuclear structure represents an
approximate implementation of Kohn-Sham density functional theory
(DFT) \cite{KS.65,Kohn,DG.90},
which is successfully employed in the treatment of the quantum
many-body problem in atomic, molecular and condensed matter physics.
At the basis of the DFT approach are energy functionals of the
ground-state density. In relativistic mean-field models,
these become functionals of the ground-state scalar density and
of the baryon current. The scalar and vector self-energies
play the role of local relativistic Kohn-Sham
potentials~\cite{SW.97,FS.00}. The mean-field models approximate
the exact energy functional, which includes all higher-order
correlations, with powers and gradients of densities,
with the truncation determined by power counting~\cite{FS.00}.  

In this work we only consider spherical even-even nuclei. Because of
time-reversal invariance the spatial
components of the four-currents vanish, and the nucleon self-energies
(reduced to the time component $\Sigma^0$ of the vector terms
$\Sigma^\mu$) read:
\begin{eqnarray}
   \Sigma^{0} & = & G_V \rho - V_C \; ,\\
   \Sigma^{0}_{T} & = & G_{TV} \rho_{3} \; ,\\
   \Sigma_S & = & G_S \rho_s + D_S \triangle \rho_s \; ,\\
   \Sigma_{TS} & = & G_{TS} \rho_{s3} \; ,\\
   \Sigma_{rS} & = & \frac{\partial D_S}{\partial \rho}
   (\nabla \rho_s ) \cdot(\nabla \rho ) \; ,\\
   \Sigma_r^{0} & = & 
   \frac{1}{2} \left\{ \frac{\partial G_S}{\partial \rho} \rho_s^2 
   + \frac{\partial G_{TS}}{\partial \rho} \rho_{s3}^2 
   + \frac{\partial G_V}{\partial \rho} \rho^2
   + \frac{\partial G_{TV}}{\partial \rho} \rho_{3}^2 
   - \frac{\partial D_S}{\partial \rho}
   (\nabla \rho_s ) \cdot (\nabla \rho_s ) \right\} \; , \nonumber \\ 
   ~ & ~ &
\end{eqnarray}
where $V_C$ is the Coulomb potential. 
%\begin{equation}
% V_C ({\bf r}) =  \frac{e^2}{4\pi} \int d{\bf x}' 
%  \frac{\rho^P({\bf r}')}{|{\bf r -r}'|} \; .
%\end{equation}  
In the {\it mean-field} approximation the isoscalar-scalar,
isoscalar-vector, isovector-scalar and isovector-vector ground state
densities are calculated from the Dirac wave functions $\psi_k$
of the occupied positive-energy orbits as:
\begin{eqnarray}
   \rho_s & = & \sum\limits_{k=1}^{A} \bar \psi_k \psi_k \; ,\\
   \rho & = & \sum\limits_{k=1}^{A} \psi^{\dagger}_k \psi_k \; ,\\
   \rho_{s3} & = & \sum\limits_{k=1}^{A} \bar \psi_k \tau_3 \psi_k \; ,\\
   \rho_{3} & = & \sum\limits_{k=1}^{A} \psi^{\dagger}_k \tau_3 \psi_k \; ,
\end{eqnarray}
respectively.
The expression for the ground state
energy of a nucleus with A nucleons reads:
\begin{eqnarray}
   E & = & \sum\limits_{k=1}^{A} \epsilon_k - \frac{1}{2} \int d^3x \Big[
   ~G_S \rho_s^2 + G_{TS} \rho_{s3}^2 + G_V \rho^2 + + G_{TV}\rho_{3}^2 
   + \rho^p V_C \nonumber\\
 & ~ & + 
   \frac{\partial G_S}{\partial \rho} \rho_s^2 \rho +
   \frac{\partial G_V}{\partial \rho} \rho^3 +
   \frac{\partial G_{TV}}{\partial \rho} \rho_3^2 \rho +
   \frac{\partial G_{TS}}{\partial \rho} \rho_{s3}^2 \rho 
   + D_S \rho_s \triangle \rho_s \nonumber\\
 & ~ & - \frac{\partial D_S}{\partial \rho} (\nabla \rho_s) \cdot
   (\nabla \rho_s) \rho + 2\frac{\partial D_S}{\partial \rho} (\nabla \rho_s)
   \cdot (\nabla \rho) \rho_s
   \Big] \; ,  
\end{eqnarray}
where $\epsilon_k$ denotes the single-nucleon energies.
In addition, for open shell nuclei pairing correlations are
included in the simple BCS approximation~\cite{GRT.90}.
After obtaining the solution of the self-consistent Dirac equation, 
the microscopic estimate for the center-of-mass correction
\begin{equation}
\label{cms}
   E_{cm} = - \frac{\langle \vec{P}_{cm}^2 \rangle}{2AM} \;,
\end{equation}
is subtracted from the total binding energy.
Here $\vec{P}^2_{cm}$ is the squared total momentum of the
nucleus with $A$ nucleons.

The details of the calculated properties of finite nuclei will, of course,
depend on the fine-tuning of the coupling parameters.
The density-dependent couplings $G_S$, $G_V$, $G_{TS}$ and $G_{TV}$ are
constrained by in-medium
QCD sum rules and chiral perturbation calculations of one- and
two-pion exchange diagrams, as described in the previous section.
For finite nuclei we must also
determine the coupling parameter of the derivative term:
$D_S$. Dimensional considerations suggest the following ansatz
\begin{equation}
   D_S(\rho) = \frac{G_S(\rho)}{{\mathcal M}^2} \; ,
\end{equation}
where ${\mathcal M}$ is a characteristic mass scale for a given
spin-isospin channel.
There is no deeper reason, however, for the parameters of the
derivative terms to have the same density dependence as the coupling
parameters of the four-fermion interaction terms. 
The simplest option, followed in this work, is to introduce only
$D_S$ and treat it as a density-independent adjustable parameter.
In this case the remaining rearrangement
contribution to the vector-self energy becomes, of course, much simplified.
As it has been emphasized by Serot and Furnstahl~\cite{FS.00b},
the empirical data set of bulk
and single-particle properties of finite nuclei can only constrain
six or seven parameters in the general expansion of the effective
Lagrangian in powers of the fields and their derivatives. In particular,
only one parameter of the derivative terms can be determined
by the binding energies and radii of spherical nuclei. 
We adjust the single remaining surface parameter $D_S$ of the
isoscalar-scalar derivative term
to properties of light and medium-heavy $N \approx Z$ nuclei.
This approximation, which was first used by Serot and
Walecka in Ref.~\cite{SW.79}, implies that the isoscalar-vector,
the isovector-scalar and the isovector-vector interactions are
considered to be purely contact interactions (no gradient terms).  

\begin{figure}
\includegraphics[scale=0.7,angle=0]{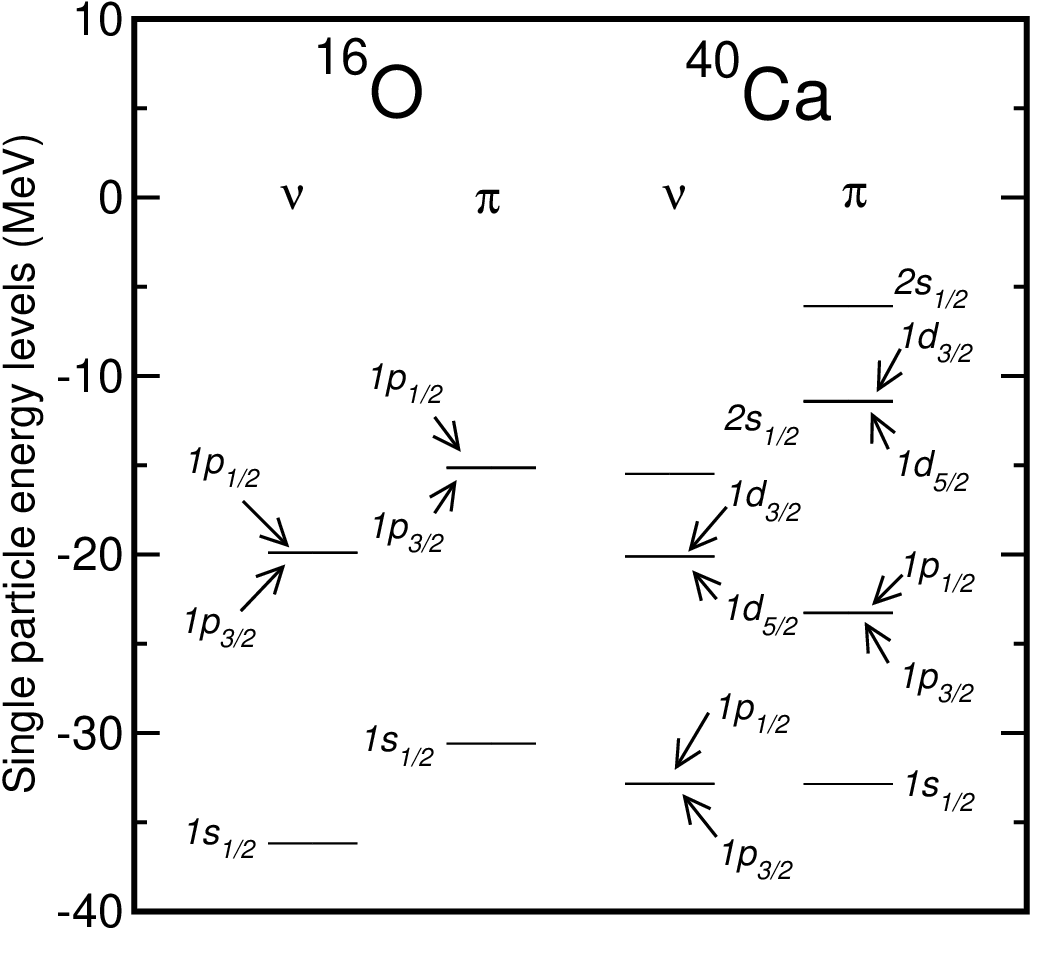}
\caption{\label{figF} Neutron and proton single-particle levels
in $^{16}$O and $^{40}$Ca calculated in the relativistic point-coupling model.
The density dependent coupling strengths include only
the contribution from chiral one- and two-pion exchange between nucleons.}
\end{figure}

Before we present the results of full model calculations
for finite nuclei, it is
instructive to consider separately the contributions of chiral
pion dynamics and condensate background self-energies
to properties of finite nuclei.
In the first step we have calculated the ground states of  
$^{16}$O and $^{40}$Ca by using the coupling parameters determined
by the nuclear matter EOS of Ref.~\cite{KFW.01} (see Fig.~\ref{figA}):
$G_S(\rho) =  G^{(\pi)}_S(\rho)$,
$G_V(\rho) =  G^{(\pi)}_V(\rho)$,
$G_{TS}(\rho) =  G^{(\pi)}_{TS}(\rho)$,
$G_{TV}(\rho) =  G^{(\pi)}_{TV}(\rho)$, $\Lambda = 646.3$ MeV,
while the couplings
to the condensate background fields are set to zero.
In this case the nuclear dynamics
is completely determined by chiral (pionic) fluctuations.
The calculated total binding energies
are already within $5-8$ \% of the experimental values, but the radii
of the two nuclei are too small (by about 0.2 fm). This is because
the spin-orbit partners $(1p_{3/2},1p_{1/2})$ and
$(1d_{5/2},1d_{3/2})$ are practically degenerate. In
Fig.~\ref{figF} we display the calculated
neutron and proton single-particle levels
of $^{16}$O and $^{40}$Ca. The energies of the degenerate doublets
are close to the empirical positions of the centroids of the
spin-orbit partner levels, and even the calculated energies
of the $s$-states are realistic. This is a very interesting result.
The chiral pion dynamics provides the saturation mechanism
and binding of nuclear matter, but not the strong
spin-orbit force. The inclusion of the isoscalar-scalar
derivative term has some effect on the calculated radii but
it cannot remove the degeneracy of the spin-orbit doublets.

The spin-orbit potential plays a central role in
nuclear structure. It is at the basis of the nuclear
shell model, and its inclusion is essential in order to
reproduce the experimentally established magic numbers.
In non-relativistic models based on the mean-field approximation,
the spin-orbit potential is included in a purely phenomenological way,
introducing the strength of the spin-orbit interaction as
an additional parameter. Its value
is usually adjusted to the experimental
spin-orbit splittings in spherical nuclei, for example $^{16}$O.
On the other hand, in the
relativistic description of the nuclear many-body problem,
the spin-orbit interaction arises naturally from the
scalar-vector Lorentz structure of the effective Lagrangian, and
relativistic models reproduce the empirical
spin-orbit splittings.
In the first order approximation, and assuming spherical
symmetry, the spin-orbit term of the effective single-nucleon
potential can be written as
\begin{equation}
\label{so1}
  V_{s.o.} = \frac{1}{2M^2} \left(
  {1 \over r}{\partial \over \partial r} V_{ls}(r)\right) {\bm l \cdot \bm s} \; ,
\end{equation}
where the large spin-orbit potential $V_{ls}$ arises 
from the difference of the vector
and scalar potentials, $V^0-S \sim 0.7 ~{\rm GeV}$ \cite{Rin.96,Bro.77}. 
Explicitly,
\begin{equation}
\label{so2}
   V_{ls} = {M \over M_{eff}} (V^0-S) \; .
\end{equation}
where
$M_{eff}$ is an effective mass specified as \cite{Rin.96}
\begin{equation}
\label{so3}
   M_{eff} = M - {1 \over 2} (V^0-S).
\end{equation}
The isoscalar nucleon self-energies generated by pion exchange
are not sufficently large to produce the empirical effective
spin-orbit potential.
The degeneracy of spin-orbit doublets is removed by
including the isoscalar condensate background nucleon self-energies
that arise through changes in the quark condensate and the quark density
at finite baryon density.

In Table~\ref{tab5} we display the
binding energies per nucleon and charge radii of light
and medium-heavy nuclei, calculated in the relativistic
point-coupling model constrained by in-medium
QCD sum rules and chiral perturbation theory, in comparison
with experimental values. In addition to the four parameters
determined by the nuclear matter equation
of state, we have adjusted the isoscalar-scalar derivative term
in the calculation of $^{16}$O and $^{40}$Ca:
$D_S = -0.713$ fm$^{4}$.
This value of $D_S$ is very close to the ones used in the effective
interactions of the standard relativistic point-coupling model
of Ref.~\cite{Burvenich:2001rh}.
It is also consistent with the ``natural'' order of magnitude expected
from $D_S \sim G_S/\Lambda^2$.
The resulting agreement between the calculated
and empirical binding energies and charge radii is indeed very good.

\begin{figure}
\includegraphics[scale=0.7,angle=0]{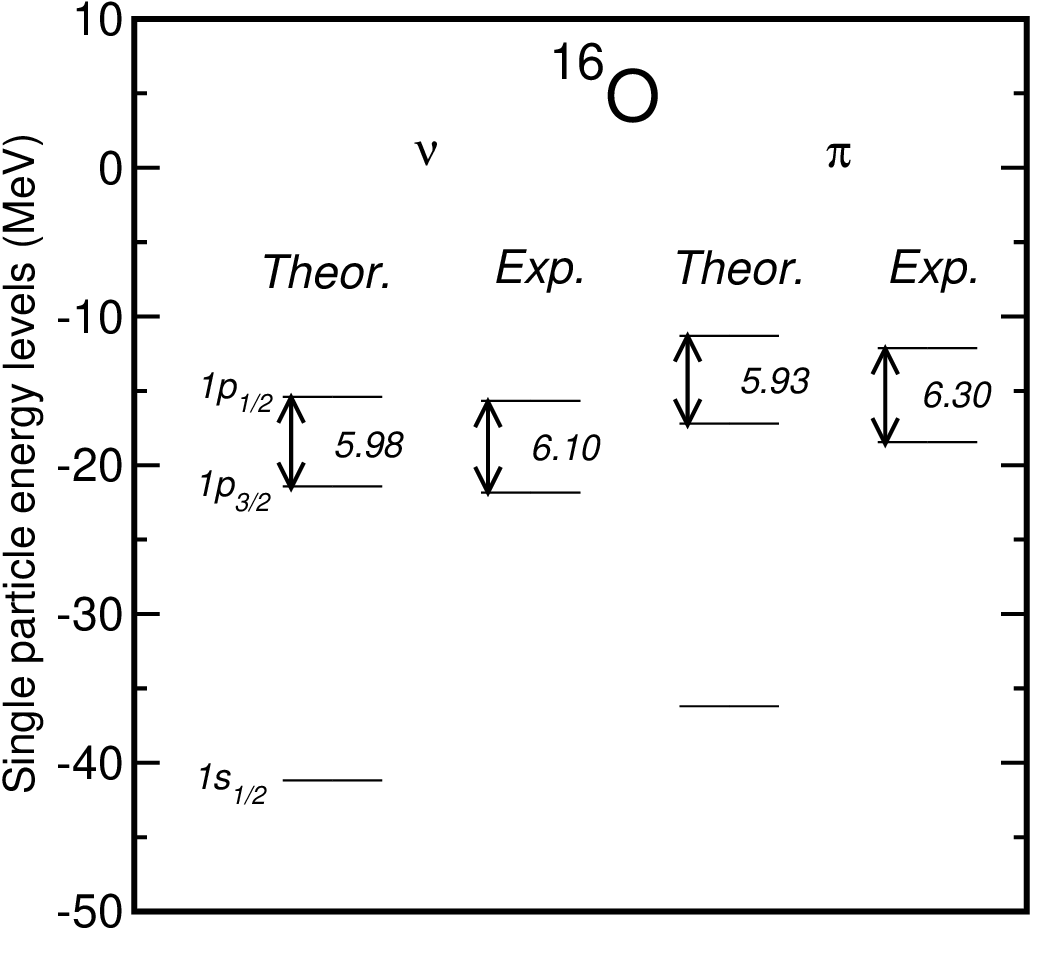}
\caption{\label{figG} The neutron and proton single-particle levels
in $^{16}$O calculated in the relativistic point-coupling model, are shown
in comparison with experimental levels. The calculation is performed
by including both the contributions of chiral pion-nucleon exchange and
of the isoscalar condensate self-energies
in the density dependent coupling strengths.}
\end{figure}

\begin{figure}
\includegraphics[scale=0.7,angle=0]{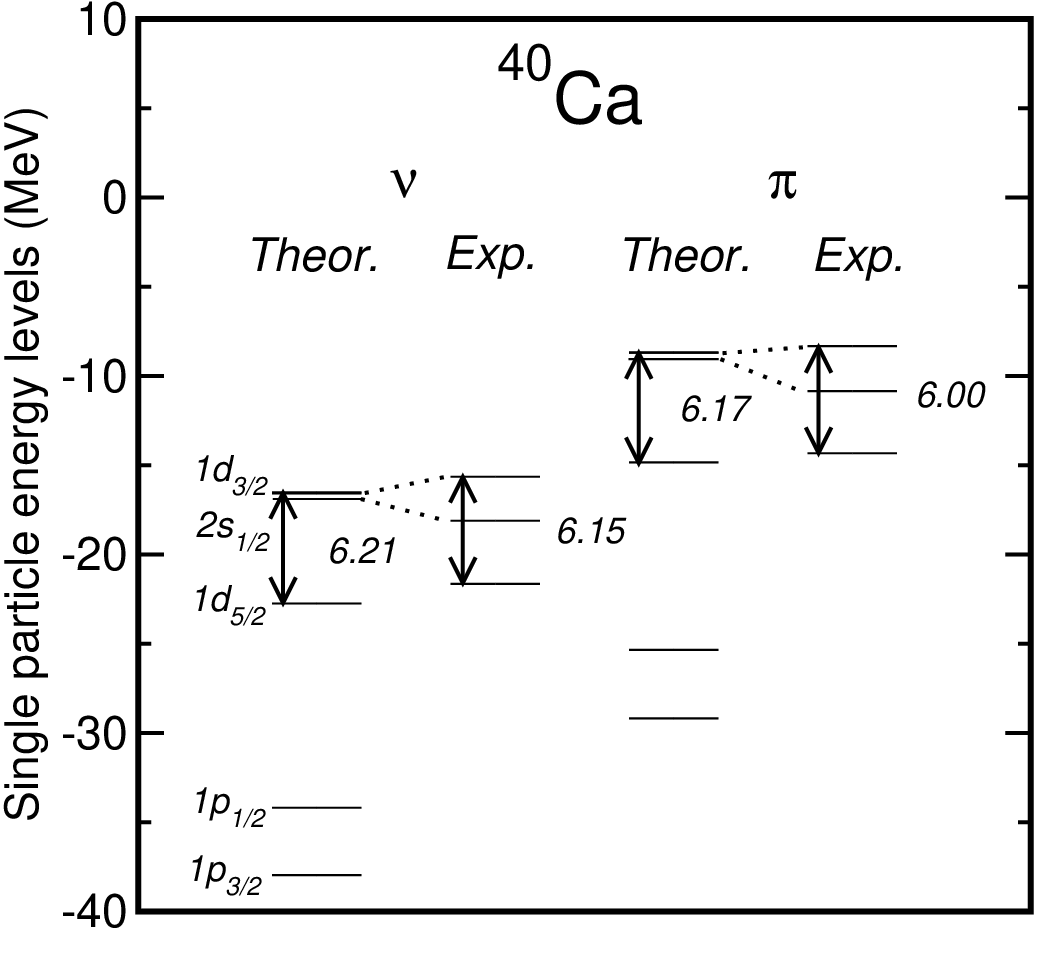}
\caption{\label{figH} Same as in Fig.~\protect\ref{figG}, but for
$^{40}$Ca.}
\end{figure}

The effect of including the isoscalar condensate background
nucleon self-energies is illustrated in Figs.~\ref{figG} and
\ref{figH}, where we display the neutron and proton single-particle
energies in $^{16}$O and $^{40}$Ca, respectively, in comparison
with the corresponding experimental levels. For both nuclei the
calculated levels are in excellent agreement with the empirical
single-nucleon levels in the vicinity of the Fermi surface.
In particular, with the inclusion of the
isoscalar condensate self-energies,
the model reproduces the empirical energy differences between
spin-orbit partner states. This is a very important result and
it shows that, while nuclear binding and
saturation are partially generated by chiral (two-pion exchange)
fluctuations in our approach, strong scalar and vector fields of 
equal magnitude and opposite sign, 
induced by changes of the QCD vacuum in the presence
of baryonic matter, generate the large effective
spin-orbit potential in finite nuclei. Not
surprisingly, the$~^{40}$Ca spectrum is reminiscent of an underlying
pseudo-spin symmetry~\cite{Gin.02}.

\begin{figure}
\includegraphics[scale=0.55,angle=0]{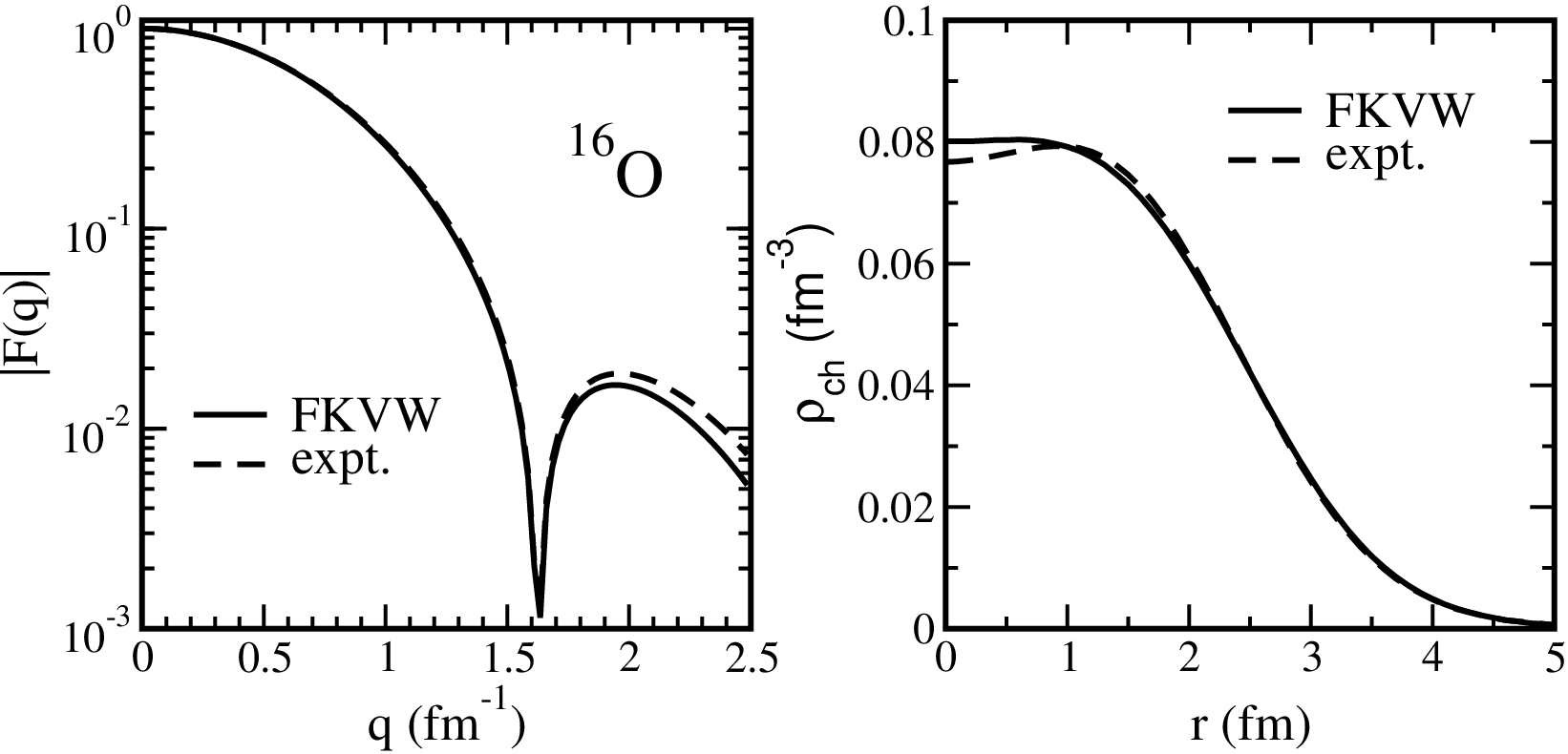}
\caption{\label{figI} 
Charge form factor (left panel) and charge density distribution 
(right panel) of $^{16}$O.  
The results obtained in the relativistic
point-coupling model, constrained by in-medium
QCD sum rules and chiral perturbation theory, are compared 
with experimental data~\protect\cite{Vri.87}.}
\vspace{1cm}
\end{figure}

\begin{figure}
\includegraphics[scale=0.55,angle=0]{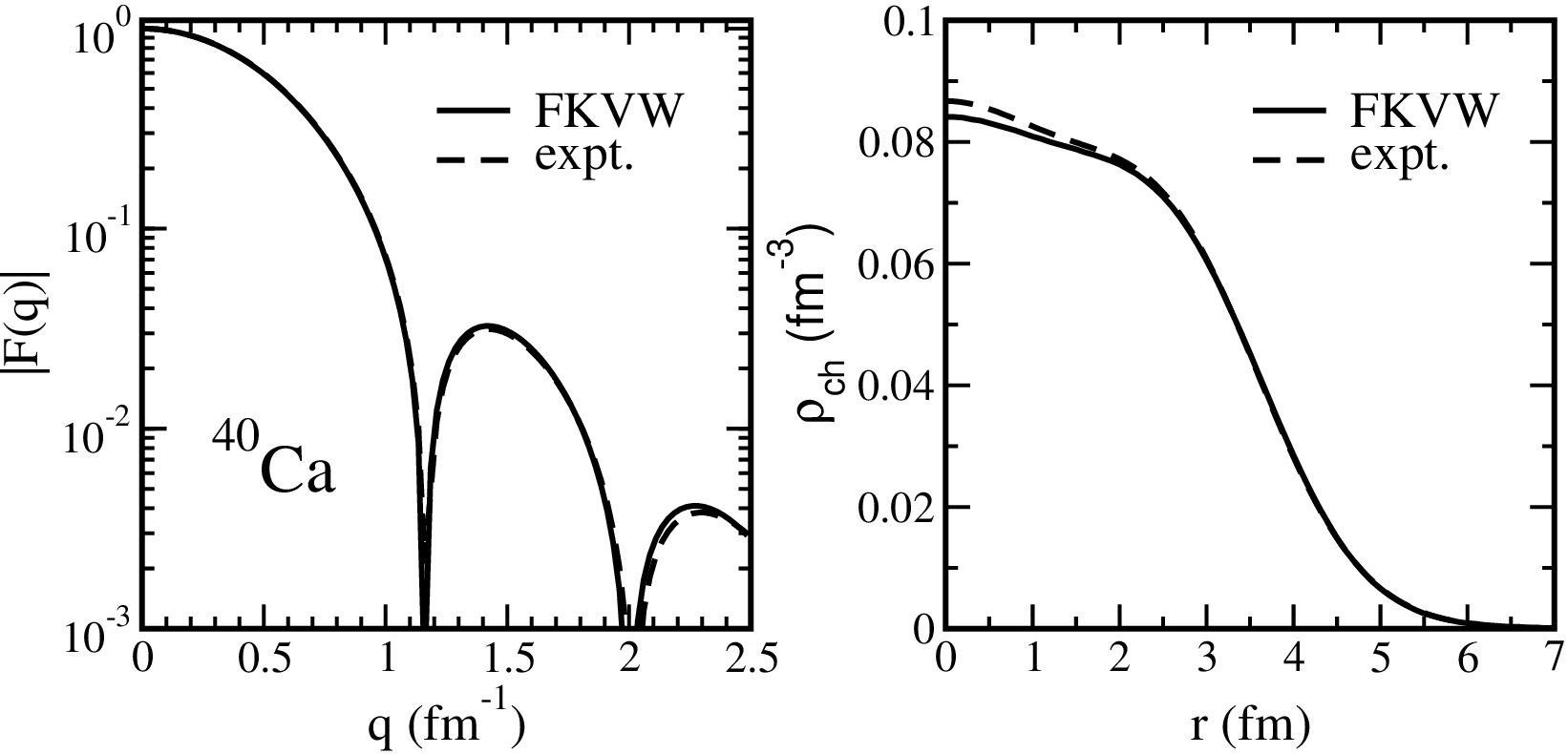}
\caption{\label{figJ} Same as in Fig.~\protect\ref{figI}, but for
$^{40}$Ca.}
\end{figure}

The charge density distributions and charge form factors
of $^{16}$O and $^{40}$Ca
are displayed in Figs. \ref{figI} and  \ref{figJ}, respectively.
The results obtained in the relativistic
point-coupling model, constrained by in-medium
QCD sum rules and chiral perturbation theory, are compared with the 
experimental charge density profiles and form factors~\cite{Vri.87}. 
The theoretical charge density distributions are obtained by folding
the point proton density distribution with the electric nucleon form factor
of dipole form. The corresponding form factors
are plotted as functions of the momentum $\bm{q}$~\cite{FST}
\begin{equation}
F_{ch}(\bm{q}) = \frac{1}{Z} \rho_{ch}(\bm{q}) \left[ 1 + 
\frac{\bm{q}^2}{8
\langle P^2_{cm}\rangle} + \ldots \right] \; ,
\end{equation}
where $\rho_{ch}(\bm{q})$ is the Fourier transform of the charge
density and the second term is a
center-of-mass correction. 
Higher-order effects in $\bm{q}^2/P^2_{cm}$ are negligible.
We notice the excellent agreement 
between the calculated and experimental charge form factors 
for momenta $|\bm{q}| \leq 2$ fm$^{-1}$. When the calculated
charge densities are compared to results obtained with 
standard relativistic mean-field
effective interactions, the present results display
less pronounced shell effects, in better agreement with empirical data.  

\begin{figure}
\includegraphics[scale=0.65,angle=0]{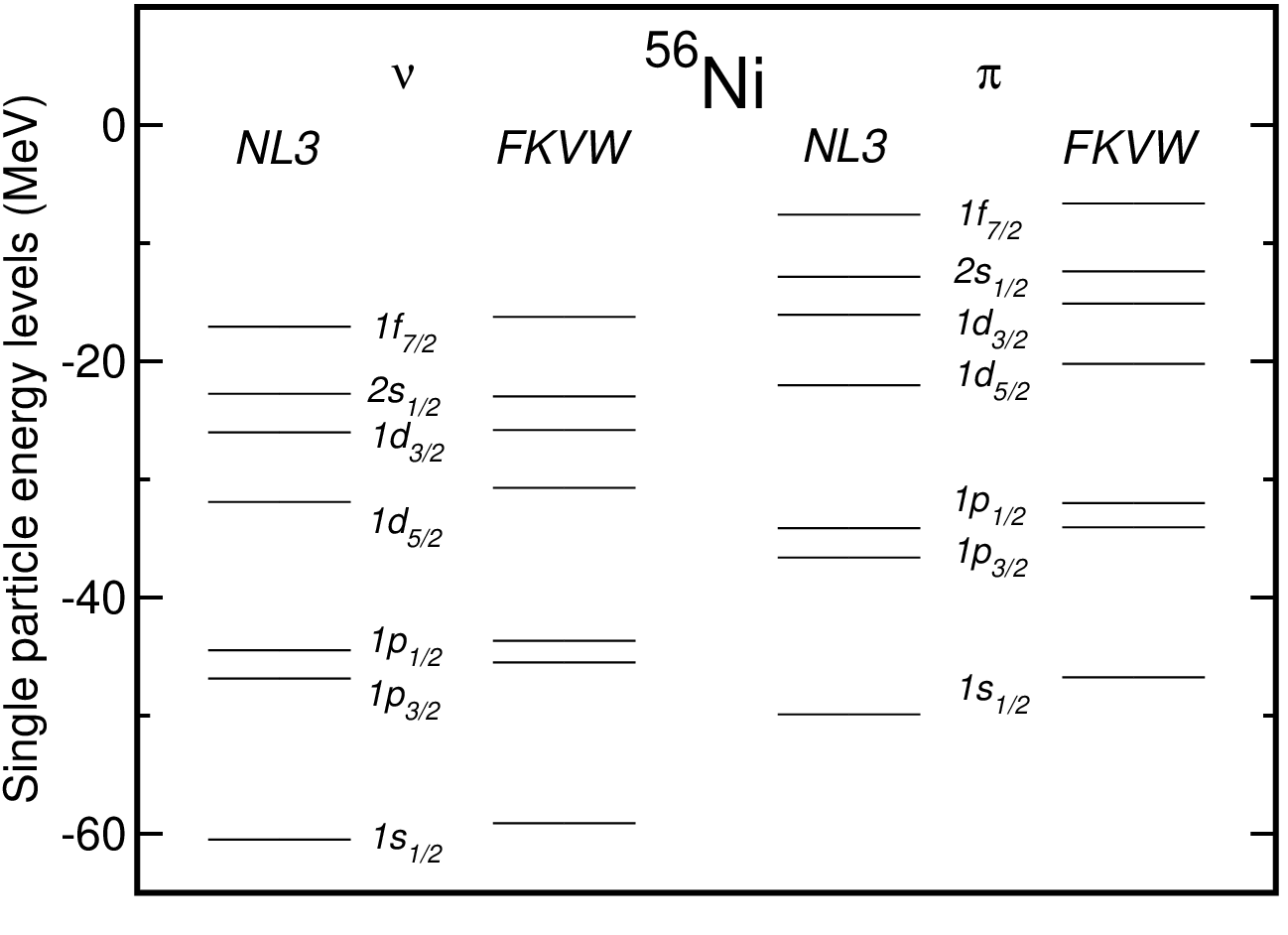}
\caption{\label{figK}Neutron and proton single-particle spectra of
$^{56}$Ni, calculated with the standard relativistic mean-field
model NL3 effective interaction~\protect\cite{LKR.97}, and
with the relativistic
point-coupling model constrained by in-medium
QCD sum rules and chiral perturbation theory (FKVW).}
\end{figure}

In Fig. \ref{figK} we  compare the single-nucleon spectra
of $^{56}$Ni, calculated in our approach,
with the results of a relativistic mean-field calculation 
using NL3~\cite{LKR.97}, probably the best non-linear 
meson-exchange effective interaction.
The spectra are in remarkable agreement. All these results
demonstrate that in the present approach, based on
QCD sum rules and in-medium chiral perturbation theory, and
with a small number of model parameters determined
directly by these constraints,
it is possible to describe symmetric and asymmetric nuclear matter,
as well as properties of finite nuclei, at a quantitative level
comparable with phenomenological relativistic mean-field models.

%=========================================================================
%  Section 5

\section{\label{secV}Summary and conclusions}
%=========================================================================

The goals of nuclear structure theory have evolved,
especially in the last decade, from the macroscopic and
microscopic descriptions of structure phenomena in stable nuclei, towards
more exotic nuclei far from the valley of $\beta$-stability, and towards the
investigation of the bridge between low-energy QCD and the
dynamics of nuclei. New experimental information
on exotic nuclei with extreme isospins present serious
challenges for traditional nuclear structure models.
It is likely that radically improved models,
or even completely new theoretical strategies and approaches, will be required
as one proceeds to these unexplored region with radioactive
nuclear beams.

Models based on effective field theories and density functional theory
provide a systematic theoretical framework for studies of
nuclear dynamics.
By using the important active degrees of freedom at low energies,
and by expanding in small parameters
determined by the relevant scales, effective Lagrangians are constructed
that provide a microscopically consistent, and yet simple and
economical approach to the nuclear many-body problem.  

In this work we have derived, following Ref.~\cite{Finelli:2002na}, a
relativistic nuclear point-coupling model
with density-dependent contact interactions,
which emphasizes the connection between nuclear
dynamics and two key features of low-energy, non-perturbative 
QCD: the presence of a non-trivial vacuum characterized by strong
condensates and the important role  of pionic fluctuations governing
the low-energy, long wavelength dynamics according to the rules imposed by
spontaneously broken chiral symmetry.

Investigations based on QCD sum rules have
shown how large isoscalar scalar and vector nucleon self-energies,
that characterize the nuclear ground state, arise in
finite-density QCD through changes in the quark
condensate and the quark density at finite baryon density.
Chiral (pionic) fluctuations superimposed on the condensate 
background fields play a prominent role for nuclear binding and saturation. 
These pionic fluctuations are calculated according to the rules of in-medium
chiral perturbation theory.

Based on these observations, we have formulated
a density-dependent point-coupling model for nuclear matter and
finite nuclei.
The coupling parameters are functions
of the nucleon density operator $\hat{\rho}$. Their functional dependence
is determined from finite density QCD sum rules and
in-medium chiral perturbation theory.
The strength
parameters of the second order isoscalar interaction terms
in the Lagrangian include
contributions from condensate background fields and
pionic (chiral) fluctuations.
For the isovector channel it is (so far) assumed that only the pionic
fluctuations contribute to the nucleon self-energies.

The resulting nuclear point-coupling model, formally used in the
mean-field approximation, but encoding effects beyond mean-field
through the density dependent couplings, has been employed in an analysis of
the equations of state for symmetric and asymmetric nuclear matter,
and of bulk and single-nucleon properties of light and
medium $N \approx Z$ nuclei. In comparison with purely
phenomenological relativistic and non-relativistic mean-field models
of nuclear structure,
the built-in QCD constraints and the explicit treatment
of one- and two-pion exchange significantly reduce the freedom in adjusting
parameters and functional forms of density-dependent
vertex functions.

A very good equation of state of symmetric nuclear
matter is obtained already by equating the isoscalar-scalar,
the isoscalar-vector, the
isovector-scalar, and the isovector-vector nucleon
self-energies in the single-nucleon Dirac equation  with
those calculated using in-medium chiral perturbation theory
up to three loop order in the energy density.
The contributions to the energy density originate
exclusively from one- and two-pion exchange between nucleons, and they are
ordered in powers of the Fermi momentum $k_f$ (modulo functions
of $k_f/m_\pi$). The empirical saturation point, the nuclear matter
incompressibility, and the asymmetry energy at saturation can be well
reproduced at order $\mathcal{O}(k_f^5)$ in the chiral expansion with just
one single momentum cut-off scale $\Lambda \simeq  0.65$ GeV which
parameterizes unresolved short-distance dynamics.

In the next step we have
included the contributions of the condensate background self-energies
in the isoscalar-scalar and isoscalar-vector channels.
We have considered corrections possibly related
to condensates of higher dimension as well as 3-body forces, 
i.e. terms of order
$\rho^2$ in the self-energies.
The isoscalar condensate background self-energies turn out to be important
also for neutron matter. However, details of the isovector interactions
(beyond chiral one- and two-pion exchange) must still be improved.

The relativistic nuclear point-coupling model,
constrained by in-medium QCD sum rules and chiral perturbation theory,
has also been tested in
self-consistent calculations of ground state properties of finite nuclei.
In the present work we have only considered spherical even-even
light and medium-heavy $N \approx Z$ nuclei.
Four parameters determine the nuclear matter
EOS: the cut-off parameter $\Lambda$ of the in-medium CHPT calculation
of two-pion exchange diagrams, the two parameters of the
isoscalar condensate background nucleon self-energies linear in the
corresponding densities: $G_{S}^{(0)}$ and $G_{V}^{(0)}$,
and the strength parameter $g_{V}^{(1)}$ of the quadratic term of the
isoscalar vector self-energy. For finite nuclei we adjust in
addition the strength parameter of the 
isoscalar-scalar derivative interaction term in the Lagrangian. This brings
the total number of adjustable parameters to five.
It has been emphasized that the empirical data set of $N \approx Z$
finite nuclei can only constrain five or
six parameters in the general expansion of an effective
Lagrangian in powers of the fields and their derivatives. In particular,
only one parameter of the derivative terms can be determined
by the binding energies and radii of spherical nuclei.
The $G_{S,V}^{(0)}$ turn out to be so remarkably 
close to leading order QCD sum rule estimates that one could have guessed
their values right from the beginning.
The remaining three parameters (the high momentum scale reflecting
unresolved short-distance physics, a ``three body'' contribution
to the nucleon self-energy and a surface (derivative) term)
behave ``naturally'' according to the power-counting hierarchy
of effective field theory. Of course, questions concerning
systematic convergence of the in-medium chiral loop expansion
still remain and need to be further explored.   

%An important result of the present
%analysis is that nuclear binding and
%saturation are essentially governed by chiral (two-pion exchange)
%fluctuations.
An important result of the present analysis
is that  chiral (two-pion exchange) fluctuations
are consistently included in the nuclear binding and 
in the saturation mechanism.
These fluctuations are superimposed on the condensate background
scalar and vector fields of equal magnitude and
opposite sign, induced by changes of the QCD vacuum in the presence
of baryonic matter. They drive the large spin-orbit splitting in finite nuclei.
The calculated neutron and proton single-particle
energies in $^{16}$O and $^{40}$Ca are 
in excellent agreement with the empirical
single-nucleon levels in the vicinity of the Fermi surface,
including the empirical energy differences between
spin-orbit partner states.

This work has demonstrated that an
approach to nuclear dynamics, constrained by
the chiral symmetry breaking pattern and the condensate structure
of low-energy QCD,
can describe symmetric and asymmetric nuclear matter,
as well as properties of finite nuclei, at a quantitative level 
comparable with phenomenological relativistic and
non-relativistic mean-field models. This is a promising
result, which opens perspectives of bridging  
the gap between basic features of
low-energy, non perturbative QCD and the rich
nuclear phenomenology.

%%%%%%%%%%%%%%%%%%%%%%%%%%%%%%%%%%%%%%%%%%%%%%%%%%%%%%%%%%%%%%%%%%%%%%%%%%%

\newpage
%%%%%%%%%%%%%%%%%%%%%%%%%%%%%%%%%%%%%%%%%%%%%%%%%%%%%%%%%%%%%%%%%%%%%%%%%%%%%%

\begin{table}
\begin{center}
\caption{The coefficients of the expansion (\protect\ref{expansion})
up to order $k_f^5$ of the in-medium CHPT isoscalar-vector,
isoscalar-scalar, isovector-vector, and isovector-scalar
nucleon self-energies.}
\bigskip
\begin{tabular}{|l|c|c|c|c|}
\hline
 ~ & {\sc V-vertex} & {\sc S-vertex}
 & {\sc TV-vertex} & {\sc TS-vertex}\\
\hline
$c_{30}$ & 3.96669 & 3.97078 & -4.00813 & -4.12669 \\
$c_{31}$ & -7.55255 & -7.55255 & 6.54555 & 6.54555 \\
$c_{32}$ & 0.762611 & 0.60693 & -0.544615 & -1.73796 \\
$c_{3L}$ & 0.29288 & 0.267627 & -0.370303 & -0.563877 \\
$c_{40}$ & 8.42137 & 8.11713 & -2.4575 & -2.97905 \\
$c_{50}$ & 10.2384 & 12.0729 & -26.8119 & -23.591 \\
$c_{5L}$ & 1.16164 & 1.92167 & -7.81327 & -6.50123\\
\hline
\end{tabular}
\label{tab1}
\end{center}
\end{table}
%%%%%%%%%%%%%%%%%%%%%%%%%%%%%%%%%%%%%%%%%%%%%%%%%%%%%%%%%%%%%%%%%%%%%%%%%%%%%%

\begin{table}
\begin{center}
\caption{The coefficients of the expansion of the in-medium
CHPT self energies (\protect\ref{prho1}) -- (\protect\ref{prho4})
in powers of the baryon density, for the value of the cut-off
parameter $\Lambda =$ 646.3 MeV. In this case we assumed
$N=Z$, neglecting isovector contributions.}
\bigskip
\begin{tabular}{|l|c|l|c|}
\hline
$c_{s1}$ & -2.805 fm$^{2}$ & $c_{ts1}$ &  1.491 fm$^{2}$\\
$c_{s2}$ & 2.738  fm$^{3}$ & $c_{ts2}$ & -1.005 fm$^{3}$\\
$c_{s3}$ & 1.346  fm$^{4}$ & $c_{ts3}$ & -1.550 fm$^{4}$\\
$c_{v1}$ & -2.718 fm$^{2}$ & $c_{tv1}$ &  2.249 fm$^{2}$\\
$c_{v2}$ & 2.841  fm$^{3}$ & $c_{tv2}$ & -0.829 fm$^{3}$\\
$c_{v3}$ & 1.325  fm$^{4}$ & $c_{tv3}$ & -1.595 fm$^{4}$\\
\hline
\end{tabular}
\label{tab2}
\end{center}
\end{table}
~
\newpage
%%%%%%%%%%%%%%%%%%%%%%%%%%%%%%%%%%%%%%%%%%%%%%%%%%%%%%%%%%%%%%%%%%%%%%%%%%%%%%%

\begin{table}
\begin{center}
\caption{Nuclear matter saturation properties: binding energy
per nucleon, saturation density and incompressibility. 
The first row corresponds to the in-medium CHPT calculation including
one- and two-pion exchange between nucleons~\cite{KFW.01}.
The EOS displayed in the second row (PC-dd) is obtained when the resulting
CHPT nucleon self-energies are mapped on the self-energies
of the relativistic point-coupling model with density dependent
couplings.}
\bigskip
\begin{tabular}{|c|c|c|c|}
\hline
   {\sc model} & E/A (MeV) & $\rho_{sat}$ ({\rm fm$^{-3}$}) & $K_0$ (MeV) \\
\hline
CHPT \protect\cite{KFW.01}& -15.26 & 0.178 & 255 \\
PC-dd & -14.51 & 0.175 & 235 \\
\hline
\end{tabular}
\label{tab3}
\end{center}
\end{table}
%%%%%%%%%%%%%%%%%%%%%%%%%%%%%%%%%%%%%%%%%%%%%%%%%%%%%%%%%%%%%%%%%%%%%%%%%%%%%%%

\begin{table}
\begin{center}
\caption{Nuclear matter saturation properties calculated
in the relativistic point-coupling model constrained by in-medium
QCD sum rules and chiral perturbation theory. In addition to the
chiral one- and two-pion exchange contribution to
the density dependence of the coupling parameters, the
nuclear matter EOS shown in the first row includes also the
isoscalar condensate background nucleon self-energies linear in the
corresponding densities. The EOS displayed in the second row
is calculated by including also the non-linear ``3-body'' contribution
(proportional to $\rho^2$) in the isoscalar vector condensate self-energy.}
\bigskip
\begin{tabular}{|c|c|c|c|c|}
\hline
 ~ & E/A (MeV) & $\rho_{sat}$ {\rm (fm$^{-3}$)} & $K_0$ (MeV) & $M^*/M$ \\
\hline
 {\sc linear} & -15.97 & 0.148 & 283 & 0.753 \\
 {\sc non-linear} & -15.76 & 0.151 & 332 & 0.620 \\
\hline
\end{tabular}
\label{tab4}
\end{center}
\end{table}
~
\newpage
%%%%%%%%%%%%%%%%%%%%%%%%%%%%%%%%%%%%%%%%%%%%%%%%%%%%%%%%%%%%%%%%%%%%%%%%%%%%%%%

\begin{table}
\begin{center}
\caption{Binding energies per nucleon and charge radii of light
and medium-heavy nuclei, calculated in the relativistic
point-coupling model constrained by in-medium
QCD sum rules and chiral perturbation theory, are compared
with experimental values.}
\bigskip
\begin{tabular}{|c|c|c|c|c|}
\hline
 ~ & $E/A^{\rm exp}~({\rm MeV})$ & $E/A~({\rm MeV})$ &
    $r_c^{\rm exp}~({\rm fm}^{-3})$
   & $r_c~({\rm fm}^{-3})$ \\
\hline
 $~^{16}{\rm O}$  & 7.976 & 8.019  & 2.730 & 2.724\\
 $~^{40}{\rm Ca}$ & 8.551 & 8.494  & 3.485 & 3.447\\
 $~^{42}{\rm Ca}$ & 8.617 & 8.482  & 3.513 & 3.450\\
 $~^{48}{\rm Ca}$ & 8.666 & 8.396  & 3.484 & 3.476\\
 $~^{42}{\rm Ti}$ & 8.260 & 8.127  & ----- & 3.536\\
 $~^{50}{\rm Ti}$ & 8.756 & 8.468  & ----- & 3.549\\
 $~^{52}{\rm Cr}$ & 8.776 & 8.494  & 3.647 & 3.614\\
 $~^{58}{\rm Ni}$ & 8.732 & 8.445  & 3.783 & 3.748\\
 $~^{64}{\rm Ni}$ & 8.777 & 8.451  & 3.868 & 3.836\\
 $~^{88}{\rm Sr}$ & 8.733 & 8.439  & 4.206 & 4.205\\
 $~^{90}{\rm Zr}$ & 8.710 & 8.443  & 4.272 & 4.252\\
\hline
\end{tabular}
\label{tab5}
\end{center}
\end{table}
~
\newpage
%%%%%%%%%%%%%%%%%%%%%%%%%%%%%%%%%%%%%%%%%%%%%%%%%%%%%%%%%%%%%%%%%%%%%%
%-------------------------------------------------------------------------
%------------------------------ -------------------------------------------
%-------------------------------------------------------------------------
%-------------------------------------------------------------------------
%-------------------------------------------------------------------------
%-------------------------------------------------------------------------
%-------------------------------------------------------------------------
%-------------------------------------------------------------------------
%-------------------------------------------------------------------------
%-------------------------------------------------------------------------
%-------------------------------------------------------------------------

\begin{thebibliography}{999}

\bibitem{SW.97} B.D. Serot and J.D. Walecka,
        Adv. Nucl. Phys. {\bf 16}, 1 (1986);
        Int. J. Mod. Phys. E{\bf 6}, 515 (1997).  
                              
\bibitem{Rin.96}  P. Ring, Progr. Part. Nucl. Phys. {\bf 37}, 193 (1996).

\bibitem{FS.00} R.J. Furnstahl and B.D. Serot,
        Nucl. Phys. A{\bf 673}, 298 (2000);
        Comments Nucl. Part. Phys. {\bf 2}, A23 (2000).

\bibitem{FST} R.J. Furnstahl, B.D. Serot, and H.-B. Tang,
        Nucl. Phys. A{\bf 615}, 441 (1997); A{\bf 640}, 505 (E) (1998).
 
\bibitem{Lin} W. Lin and B.D. Serot, Phys. Lett. B{\bf 233}, 23 (1989);
        Nucl.  Phys. A{\bf 512}, 637 (1990).

\bibitem{FS.00b} R.J. Furnstahl and B.D. Serot,
        Nucl. Phys. A{\bf 671}, 447 (2000). 
             
\bibitem{Man83} A.~Manohar and H.~Georgi,
        Nucl. Phys. B{\bf 234}, 189 (1984).

\bibitem{FML.96} J.L. Friar, D.G. Madland and B.W. Lynn,
        Phys. Rev. C{\bf 53}, 3085 (1996).
        
\bibitem{Rusnak:1997dj} J.~J.~Rusnak and R.~J.~Furnstahl,
        Nucl.\ Phys.\ A{\bf 627}, 495 (1997).

\bibitem{Burvenich:2001rh} T.~Burvenich, D.~G.~Madland,
        J.~A.~Maruhn and P.~G.~Reinhard,
        Phys. Rev. C{\bf 65}, 044308 (2001).

\bibitem{CFG.91} T.D. Cohen, R.J. Furnstahl and D.K. Griegel,
        Phys. Rev. Lett. {\bf 67}, 961 (1991);
        Phys. Rev. C{\bf 45}, 1881 (1992).
        
\bibitem{DL.90} E.G. Drukarev and E.M. Levin, Nucl. Phys. A{\bf 511}, 679
        (1990); Prog. Part. Nucl. Phys. {\bf 27}, 77 (1991).
        
\bibitem{Jin.94} X. Jin, M. Nielsen, T.D. Cohen, R.J. Furnstahl
        and D.K. Griegel, Phys. Rev. C{\bf 49}, 464 (1994).

\bibitem{Ham_Fur} H.~W.~Hammer and R.~J.~Furnstahl,
        Nucl. Phys. A{\bf 678}, 277 (2000).

\bibitem{KFW.01} N. Kaiser, S. Fritsch, and W. Weise,
        Nucl. Phys. A{\bf 697}, 255 (2002).

\bibitem{KFW.02} N. Kaiser, S. Fritsch and W. Weise,
        Nucl. Phys. A{\bf 700}, 343 (2002).

\bibitem{Finelli:2002na} P.~Finelli, N.~Kaiser, D.~Vretenar and W.~Weise,
         Eur. Phys. J. A{\bf 17}, 573 (2003).

\bibitem{FTS} R.J. Furnstahl, H.-B. Tang, and B.D. Serot,
        Phys. Rev. C{\bf 52}, 1368 (1995).

\bibitem{FLW.95} C. Fuchs, H. Lenske, and H.H. Wolter,
        Phys. Rev. C{\bf 52}, 3043 (1995).

\bibitem{Manakos:wu} P.~Manakos and T.~Mannel,
        Z.\ Phys.\ A{\bf 334}, 481 (1989).

\bibitem{Mad} D.G. Madland, B.A. Nikolaus and T. Hoch,
        Phys. Rev. C{\bf 46}, 1757 (1992).
        
\bibitem{JL.98} F. de Jong and H. Lenske,
        Phys. Rev. C{\bf 57}, 3099 (1998).
                
\bibitem{HKL.01} F. Hofmann, C.M. Keil and H. Lenske,
        Phys. Rev. C{\bf 64}, 034314 (2001).

\bibitem{TW.99} S. Typel and H.H. Wolter,
        Nucl. Phys. A{\bf 656}, 331 (1999).       

\bibitem{Niksic:2002yp} T.~Niksic, D.~Vretenar, P.~Finelli and P.~Ring,
        Phys.\ Rev.\ C{\bf 66}, 024306 (2002).
        
\bibitem{Pic} A. Pich and J. Prades, Nucl. Phys. Proc.
        Suppl. {\bf 86}, 236 (2000); B. L. Ioffe,
        Phys. At. Nucl. (Yad. Fiz.) {\bf 66}, 30 (2003), and refs. therein.

\bibitem{Iof} B. L. Ioffe, Nucl. Phys. B{\bf 188}, 317 (1981).

\bibitem{Gas} J. Gasser, H. Leutwyler and M. Sarinio,
        Phys. Lett. B{\bf 253}, 252 (1991).

\bibitem{FP.81} B. Friedman and V.R. Pandharipande,
        Nucl. Phys. A{\bf 361}, 502 (1981).

\bibitem{Vre.97} D. Vretenar, G.A. Lalazissis, R. Behnsch,
        W. P\" oschl and P. Ring, Nucl. Phys. A{\bf 621}, 853 (1997).
      
\bibitem{Ma.01} Zhong-yu Ma, Nguyen Van Giai, A. Wandelt,
        D. Vretenar and P. Ring,
        Nucl. Phys. A{\bf 686}, 173 (2001).
    
\bibitem{Lee.98} C.-H. Lee, T.T.S. Kuo, G.Q. Li and G.E. Brown,
        Phys. Rev. C{\bf 57}, 3488 (1998).
    
\bibitem{Fur.01} R.J. Furnstahl, Nucl. Phys. A{\bf 706}, 85 (2002).

\bibitem{SH.94} V.E. Starodubsky and N.M. Hintz,
        Phys. Rev. C{\bf 49}, 2118 (1994).
    
\bibitem{Kra.94} A. Krasznahorkay {\it et al}.,
        Nucl. Phys. A{\bf 567}, 521 (1994).
    
\bibitem{ZBL.99} W. Zuo, I. Bombaci and U. Lombardo,
        Phys. Rev. C{\bf 60}, 024605 (1999).

\bibitem{Ste.00} J.~V.~Steele, [arXiv:nucl-th/0010066].

\bibitem{Gro.98} T.~Gross-Boelting, C.~Fuchs and A.~Faessler,
        Nucl.\ Phys.\ A{\bf 648}, 105 (1999).
        
\bibitem{Bog.03} S.~K.~Bogner, T.~T.~Kuo and A.~Schwenk,
        Phys.\ Rept.\  {\bf 386}, 1 (2003);
        A.~Schwenk, G.~E.~Brown and B.~Friman,
        Nucl.\ Phys.\ A{\bf 703}, 745 (2002).

\bibitem{KS.65} W. Kohn and L. J. Sham, Phys.\ Rev.\ A{\bf 140}, 1133 (1965).
  
\bibitem{Kohn} W. Kohn,  Rev.\ Mod.\ Phys.\
        {\bf 71}, 1253 (1999).
 
\bibitem{DG.90}R. M. Dreizler and E. K. U. Gross,
        {\it Density Functional Theory}, (Springer, 1990).

\bibitem{GRT.90} Y.K. Gambhir, P. Ring and A. Thimet, {\it Ann. Phys. (N.Y.)}
        {\bf 198}, 132 (1990).

\bibitem{SW.79} B.D. Serot and J.D. Walecka,
        Phys. Lett. B{\bf 87}, 172 (1979).

\bibitem{Bro.77} R. Brockmann and W. Weise, Phys. Rev. C{\bf 16}, 677 (1977).

\bibitem{Gin.02} J.~N.~Ginocchio,
        Phys.\ Rev.\ C{\bf 66}, 064312 (2002).

\bibitem{Vri.87} H. de Vries, C.W. de Jager, and C. de Vries,
        At. Data Nucl. Data Tables {\bf 36}, 495 (1987).

\bibitem{LKR.97} G.A. Lalazissis, J. K\"onig, and P. Ring,
        Phys. Rev. C{\bf 55}, 540 (1997).

\end{thebibliography}
\end{document}